\documentclass[fleqn,usenatbib]{mnras}

\usepackage{newtxtext,newtxmath}
\usepackage[T1]{fontenc}

\DeclareRobustCommand{\VAN}[3]{#2}
\let\VANthebibliography\thebibliography
\def\thebibliography{\DeclareRobustCommand{\VAN}[3]{##3}\VANthebibliography}

\usepackage{graphicx}	
\usepackage{ulem} 
\usepackage{amsmath}	
\newcommand{\beq}{\begin{equation}}
\newcommand{\beqa}{\begin{eqnarray}}
\newcommand{\eeq}{\end{equation}}
\newcommand{\eeqa}{\end{eqnarray}}
\usepackage{xcolor}

\title[Energy distribution and substructure formation ...]{Energy distribution and substructure formation in astrophysical MHD simulations}

\author[F. Kayanikhoo et al.]{
 Fatemeh Kayanikhoo,$^{1}$\thanks{E-mail: fatima@camk.edu.pl}
Miljenko \v{C}emelji\'{c},$^{1,2,3}$
Maciek Wielgus$^{4,2}$
and Włodek Klu{\'z}niak$^{1}$
\\
$^{1}$Nicolaus Copernicus Astronomical Center of the Polish Academy of Sciences, Bartycka 18, 00-716 Warsaw, Poland \\
$^{2}$Research Centre for Computational Physics and Data Processing, Institute of Physics, Silesian University
in Opava,\\ Bezru\v{c}ovo n\'am.~13, CZ-746\,01 Opava,
Czech Republic\\
$^{3}$Academia Sinica, Institute of Astronomy and Astrophysics, P.O. Box 23-141,
Taipei 106, Taiwan\\
$^{4}$Max-Planck-Institut f\"ur Radioastronomie, Auf dem H\"ugel 69, D-53121 Bonn, Germany
}

\date{Accepted XXX. Received YYY; in original form ZZZ}

\pubyear{2023}

\begin{document}
\label{firstpage}
\pagerange{\pageref{firstpage}--\pageref{lastpage}}
\maketitle

\begin{abstract}
During substructure formation in magnetized astrophysical plasma, dissipation of magnetic energy facilitated by magnetic reconnection affects the system dynamics by heating and accelerating the ejected plasmoids. Numerical simulations are a crucial tool for investigating such systems. 
In astrophysical simulations, the energy dissipation,
reconnection rate and substructure formation critically depend on the onset of reconnection of numerical or physical origin.
In this paper, we hope to assess the reliability of the state-of-the-art numerical codes, PLUTO and KORAL by quantifying and discussing the impact of dimensionality, resolution, and code accuracy on magnetic energy dissipation, reconnection rate, and substructure formation.
We quantitatively compare results obtained with relativistic and non-relativistic, resistive and non-resistive, as well as two- and three-dimensional setups performing the Orszag-Tang test problem. 
We find sufficient resolution in each model, for which numerical error is negligible and the resolution does not significantly affect the magnetic energy dissipation and reconnection rate. The non-relativistic simulations show that at sufficient resolution, magnetic and kinetic energies convert to internal energy and heat the plasma. In the relativistic system, energy components undergo mutual conversion during the simulation time, which leads to a substantial increase in magnetic energy at 20\% and 90\% of the total simulation time of $10$ light-crossing times---the magnetic field is amplified by a factor of five due to relativistic shocks. We also show that the reconnection rate in all our simulations is higher than $0.1$, indicating plasmoid-mediated regime. It is shown that in KORAL simulations more substructures are captured than in PLUTO simulations.
\end{abstract}


\maketitle
\section{Introduction}

Dissipation processes in astrophysical plasma, including magnetic reconnection \citep{Mag_rec}, are of fundamental relevance for our understanding of a variety of observed systems, such as solar flares \citep{solar1946, Jiang_2021} or magnetic sub-storms in the Earth's magnetosphere \citep{Akasofu1968, Robert1979}. The relative motion in plasmas and gas often leads to the formation of shocks.
 Non-relativistic magnetized shocks in supernovae remnants are possible sources of acceleration of cosmic rays \citep{MHDshock, astroshock, CRshock, non-relshocks}. Energy dissipation in the relativistic regime leads to spectacular displays, such as jets and relativistic ejections from the accretion systems around compact objects \citep{osti_Jet, Ripperda2022}, or event horizon scale synchrotron emission \citep{Kinetic} and flaring \citep{Dexter2020,Wielgus2022} in the hot advection-dominated accretion flows. In the context of accretion onto compact objects, understanding dissipation occurring on small spatial scales is crucial to finding realistic sub-grid physics prescriptions for global simulations.

 Magnetic reconnection is a process by which the magnetic field lines in a plasma break and reconnect, releasing stored energy in the form of heat, particles/plasmoid acceleration, or radiation. Reconnection often occurs spontaneously and is usually associated with the presence of a current sheet, a region where the magnetic field lines become almost antiparallel and the plasma conductivity is finite. The magnetic field lines can break and reconnect due to the tearing instability, which is driven by the pressure of the plasma and the tension of the magnetic field \citep{Coppi66, Komiss07,delZa16}. Spontaneous reconnection is relatively slow, and the rate is determined by the local plasma conditions \citep{Sweet1958, Baty_2000}. Petschek proposed a shock geometry that allows fast magnetic reconnection to occur \citep{Petscheck1964}, this may be realized in magnetohydrodynamic (MHD) simulations for large values of (anomalous) resistivity. In systems with the strongly magnetized plasma, \cite{Lazarian99} state that the reconnection will always occur at some upper limit of the reconnection rate. Another scenario is the forced magnetic reconnection, which occurs due to external perturbation in a turbulent system \citep{Vekstein1998, Potter2019, Srivastava_2019}. In this scenario, the reconnection rate can be much faster than spontaneous reconnection, as the external forces can overcome the moderating resistances of the plasma. Such turbulent systems can be found in various environments, such as solar wind, the interstellar medium, or the accretion disks around black holes and neutron stars.

In this work, we study energy dissipation and magnetic reconnection in the MHD framework, using a simple example of a vortical system, the Orszag-Tang (OT) vortex \citep{small-scaleorszag_tang_1979}, a popular test problem for numerical MHD codes. In such a system, the magnetic field lines stretch and twist thus facilitating the reconnection process. This test has already been performed with state-of-the-art codes like Athena++ \citep{Athena++}, BHAC \citep{BHAC_porth}, and HARM \citep{harm2003}. Here, we implement the OT test in two more state-of-the-art codes used in numerical simulations of accretion.  We quantitatively compare results obtained with the two codes of our choice at different resolutions and setups in relativistic/non-relativistic, resistive/non-resistive, and two-dimensional (2D) vs. three-dimensional (3D) configurations, to study how much these different aspects impact the obtained results, characterized by the energy balance and reconnection rate. 

The well-established codes we selected for the comparison are the widely used, public PLUTO code \citep{m07} and the radiative, general relativistic code KORAL \citep{Sadowski_2013, koral01}.

PLUTO has extensively been used in simulations of magnetospheric star-disk interaction with alpha-viscous disk in \cite{ZF09,Cem19}, with magneto-rotational instability including alpha-dynamo in \cite{Flock11}, jet launching disks in \cite{Tzeferakos09}, accretion-ejection problem in \cite{Stepanovs14}, to mention only some. It was also used in the simulations of star-planet magnetospheric interaction, e.g. in \cite{Str14} and \cite{Varela18} and related
papers. A radiative module was included in simulations of accretion columns in classical T Tauri stars in \cite{Colombo19}.
KORAL code is used to study the accreting compact objects in general relativity involving radiation using M1 closure scheme \citep{Sadowski_2013}. The code has been used to study the radiative black hole accretion discs \citep{sadowski2014, Sadowski2017, Lancova2019,Chael2019} as well as 
super-Eddington accretion onto magnetized neutron stars \citep{Abarca_2021}.

The paper is organized as follows: in \S\ref{formalism} we review the theoretical framework, including the formalism of the MHD equations. The initial conditions in the OT problem in 2D and 3D setups are given in \S\ref{inb}. In \S\ref{resdis} we discuss the results in different cases. The reconnection rate is studied in \S\ref{vr}. In \S\ref{Scodes} we present the direct comparison of the results in the two codes we used here and we conclude in \S\ref{concl}. 

\section{Special relativistic resistive MHD equations}\label{formalism}
We investigate the energy distribution in astrophysical simulations in
the following setups:
\begin{itemize}
\item[$\bullet$] Ideal nonrelativistic magnetohydrodynamics (Ideal-MHD), 
\item[$\bullet$] Ideal relativistic MHD (Rel-MHD),
\item[$\bullet$] Resistive nonrelativistic MHD (Res-MHD).
\end{itemize}
We begin with presenting the resistive special relativistic MHD equations in Minkowski spacetime, which we then simplify to relativistic ideal MHD and non-relativistic resistive MHD cases.
The simulations are performed in the PLUTO and KORAL codes, with the exception of Res-MHD, which is performed in PLUTO alone (KORAL only treats non-resistive MHD equations).

The dynamics of magnetic fluids can be described using the equations of conservation of mass, momentum, and energy, as well as the Maxwell-Faraday, Amp\`ere-Maxwell, and Ohm equations.
For a fluid propagating in the laboratory reference frame with bulk velocity
$\boldsymbol{\upsilon}=\boldsymbol{\beta} c$,
the  Lorentz factor is defined as $\Gamma=(1-\beta^2)^{-1/2}$, and the fluid four-velocity is $\boldsymbol{u} = (\Gamma c, \Gamma \boldsymbol{\upsilon} )$. We denote fluid rest mass density in the fluid frame by $\rho c^2$, fluid pressure by $p$, fluid internal energy density in the fluid frame by $U_{\rm int}$, electric field by $\boldsymbol{E}$, and magnetic field by $\boldsymbol{B}$. The $\boldsymbol{E}$ and $\boldsymbol{B}$ fields were redefined to absorb a factor of $1/\sqrt{4\pi}$ each, so that factors of $1/(4\pi)$ do not appear in relations such as Eqs.~\ref{relmomentum} 
and \ref{en-density}, \ref{elmagtensor}.
Furthermore, we define enthalpy density in the fluid frame,
\beqa
\omega = \rho c^2 + U_{\rm int} + p \ ,
\label{eq:enthalpy}
\eeqa
momentum density 
\beqa
    \boldsymbol{m}=\omega \Gamma^2 \boldsymbol{\upsilon}+ c\boldsymbol{E}\times\boldsymbol{B} \ , 
    \label{relmomentum}
\eeqa
and the total energy density $\varepsilon$
\beqa
   \varepsilon=\omega \Gamma ^2 -p + \frac{1}{2}(E^2+B^2) \ .
   \label{en-density}
\eeqa
The conservation equations are then
\beqa
    \frac{\partial (\Gamma \rho) }{\partial t}+\boldsymbol{\nabla} \cdot ( \Gamma \rho \boldsymbol{\upsilon})=0,
\label{mass}\\
    \frac{\partial \boldsymbol{m}}{\partial t}+\boldsymbol{\nabla} \cdot \left[\omega \Gamma^2 \boldsymbol{\upsilon} \boldsymbol{\upsilon} + c^2(p \boldsymbol{I} + \boldsymbol{T}_{EM})\right]=0,
    \label{energy-momentum1}\\
    \frac{\partial {\varepsilon}}{\partial t}+\boldsymbol{\nabla} \cdot ( \omega \Gamma^2 \boldsymbol{\upsilon}+ c\,\boldsymbol{E}\times\boldsymbol{B} )=0,
    \label{energy-momentum2}
\eeqa
where additionally we denote identity matrix with $\boldsymbol{I}$, and the electromagnetic stress tensor with $\boldsymbol{T}_{EM}$, hence
\beqa
\boldsymbol{T}_{EM} = \frac{1}{2}(E^2+B^2)\boldsymbol{I} - (\boldsymbol{E}\boldsymbol{E} + \boldsymbol{B}\boldsymbol{B})\ .
\label{elmagtensor}
\eeqa

The Maxwell-Faraday and Amp\`ere-Maxwell equations are
\beqa
     &\frac{1}{c} \frac{\partial \boldsymbol{B}}{\partial t}+\boldsymbol{\nabla}\times \boldsymbol{E}=0, \label{EM1}\\
    & \frac{1}{c}\frac{\partial \boldsymbol{E}}{\partial t}-\boldsymbol{\nabla}\times\boldsymbol{B}=-\boldsymbol{J}/c, \label{EM2} 
\eeqa
respectively, where $\boldsymbol{J}$ is the current density that comes from Ohm's law, 

\beqa
   \boldsymbol{J}=({\Gamma c^2}/{\eta})(\boldsymbol{E}+\boldsymbol{\beta}\times
   \boldsymbol{B}),
\label{eq:currents_general}    
\eeqa
%
%
%
where $\eta$ is the magnetic diffusivity, which is identical to resistivity. The additional condition $\nabla\cdot~\boldsymbol{B}=0$ from Gauss's law is enforced during the numerical evolution of the magnetic field.

In order to obtain the system of nonrelativistic resistive MHD equations from Eqs. \ref{mass}-\ref{energy-momentum2}, we make a number of approximations based on $\beta \ll 1$ and $p+U_{\rm int} \ll \rho$ assumptions, leading to a following formulation:
\beqa
    \frac{\partial \rho }{\partial t}+\boldsymbol{\nabla} \cdot (\rho \boldsymbol{\upsilon})=0,
\label{mass_nonrel}\\
    \frac{\partial  \rho \boldsymbol{\upsilon}}{\partial t}+\boldsymbol{\nabla} \cdot (\rho  \boldsymbol{\upsilon} \boldsymbol{\upsilon} +p\boldsymbol{I} +\boldsymbol{T}_{EM})=0,
\label{energy-momentum1_nonrel}\\
    \frac{\partial {\varepsilon}}{\partial t}+\boldsymbol{\nabla} \cdot \left[ \left( \omega + \rho \frac{\upsilon^2}{2} \right) \boldsymbol{\upsilon} + c\boldsymbol{E}\times\boldsymbol{B} \right]=0,
\label{energy-momentum2_nonrel}
\eeqa
where the non-relativistic total energy  
and enthalpy densities are 
\beqa
   \varepsilon= U_{\rm int}  + \rho \frac{\upsilon^2}{2} + \frac{1}{2}(E^2+B^2) \ ,
   \label{en-density_nonrel}
\eeqa

\beqa
\omega = U_{\rm int} + p \ .
\label{eq:enthalpy_nonrel}
\eeqa
Additionally, Ohm's law in resistive nonrelativistic MHD becomes

\beqa
    \boldsymbol{J} = \frac{c^2}{\eta}(\boldsymbol{E}+\boldsymbol{\beta}\times \boldsymbol{B}) = c \boldsymbol{\nabla} \times \boldsymbol{B} \ ,
    \label{current_nonrel}
\eeqa
neglecting the displacement currents ($\partial \boldsymbol{E}/ \partial t = 0$) in Eq.~\ref{EM2} to obtain the second equality.

The diffusive time scale, $\tau_{\eta}=L^2/ \eta$ (in conventional units, if $\eta$ is in $\mathrm{cm^2/s}$ and $L$ is in  $\mathrm{cm}$, then $\tau_{\eta}$ is in seconds) can be
compared with the dynamical time scale $\tau_{\upsilon}=L/\upsilon$, where $L$ is the characteristic length scale of the system
and $\upsilon$ is the characteristic velocity scale. The ratio of the two time scales is known as the magnetic Reynolds number
\beqa
    R_{\rm m} = \frac{\tau _{\eta}}{\tau _\upsilon} = \frac{\upsilon L}{\eta}  \ .
\eeqa
When the typical velocity scale of the system is the Alfv\'{e}n velocity $\upsilon_{\rm A}=B/\sqrt{4\pi\rho}$, this ratio is called the Lundquist number 
\beqa
S=\upsilon_A L/\eta.
\eeqa
Astrophysical systems often satisfy the condition $S \gg 1$, which is equivalent to $L \gg \eta /\upsilon_A$. In such cases, for either relativistic or non-relativistic cases, we can use the ideal MHD approximation
\beqa
\boldsymbol{E} = \boldsymbol{B} \times \boldsymbol{\beta}.
\label{idealfield}
\eeqa
As a consequence, $\boldsymbol{E}$ can be readily evaluated and does not need to be evolved with the Amp\`ere-Maxwell equation (Eq. \ref{EM2}), simplifying the Maxwell-Faraday equation (Eq. \ref{EM1}) for the $\boldsymbol{B}$ field evolution to
\beqa
    \frac{1}{c}\frac{\partial \boldsymbol{B}}{\partial t} + \boldsymbol{\nabla}\times \left( \boldsymbol{B} \times \boldsymbol{\beta} \right) = 0 \ . 
\eeqa

\section{Orszag-Tang test problem}\label{inb}
\begin{figure*}
\centering
    \includegraphics[width=1\columnwidth]{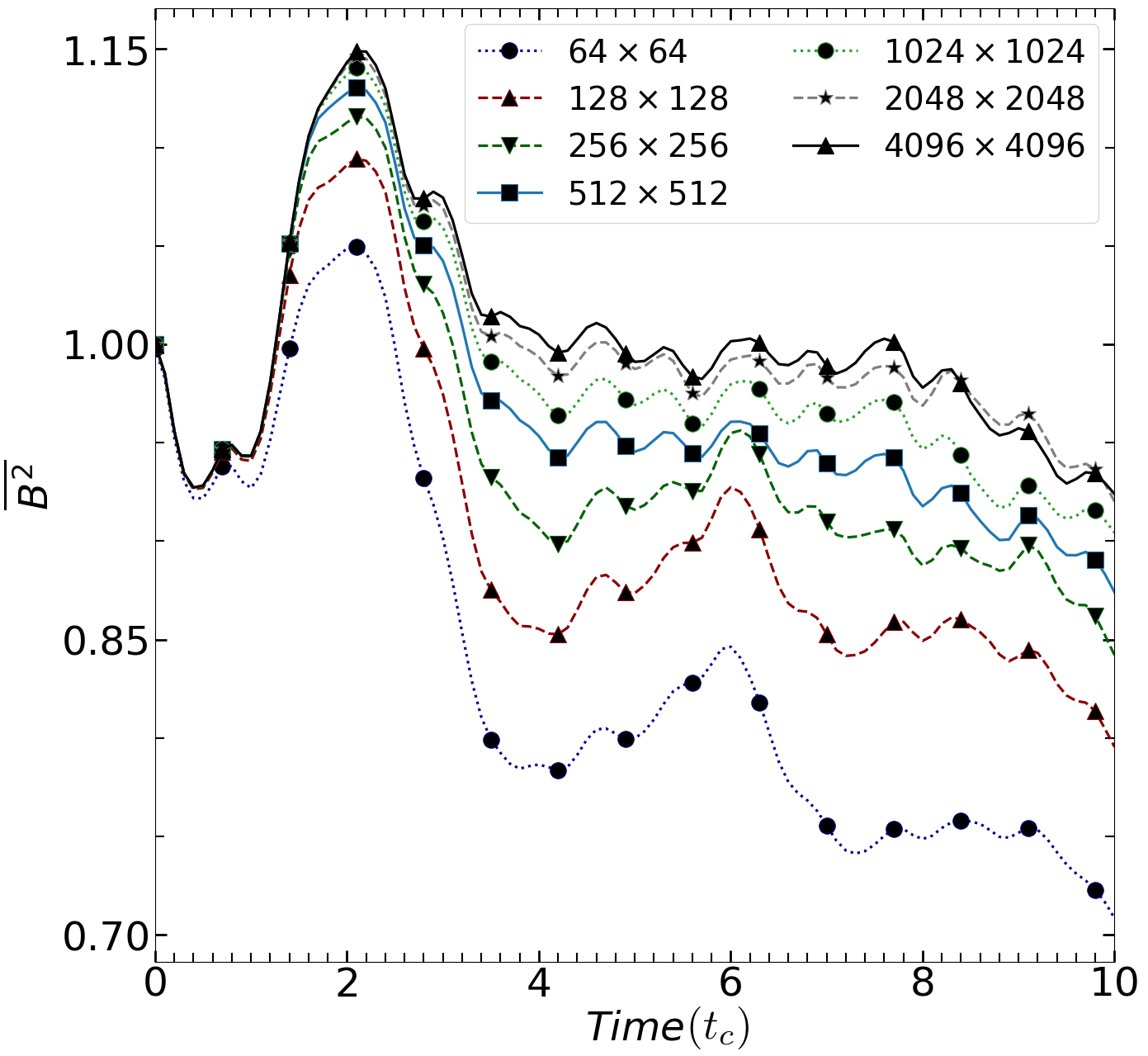}
    \includegraphics[width=1\columnwidth]{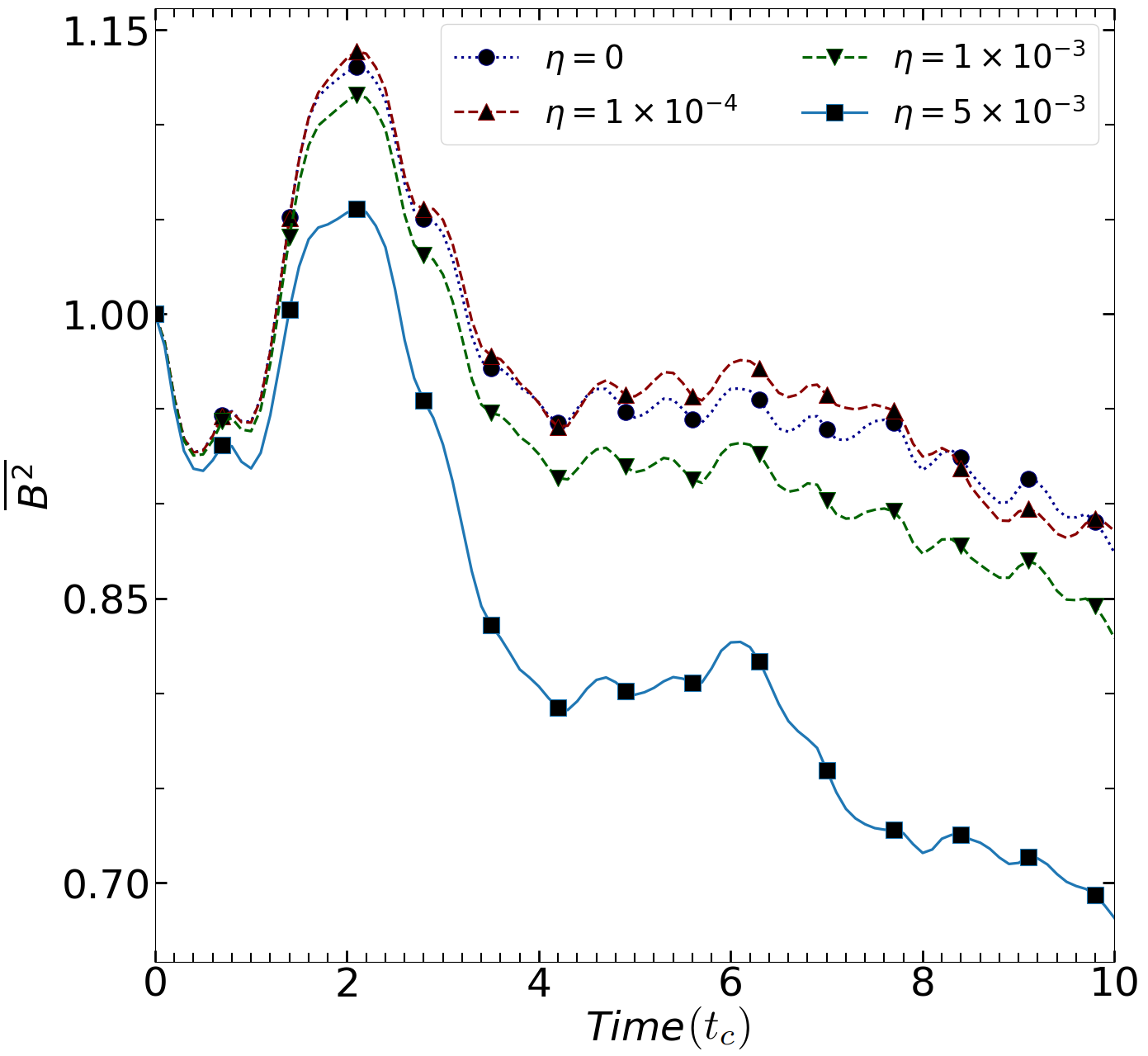}
    \caption{The time evolution of $\overline{B^2}$ in PLUTO simulations: the Ideal-MHD case with different resolutions ({\it left panel}), and Res-MHD case with different physical resistivities $\eta$ for the resolution of $512^2$ ({\it right panel}). The unit of time is $t_c=L/c$.}
\label{MHD/Res}
\end{figure*}

\begin{figure*}
\centering
\includegraphics[width=1\columnwidth]{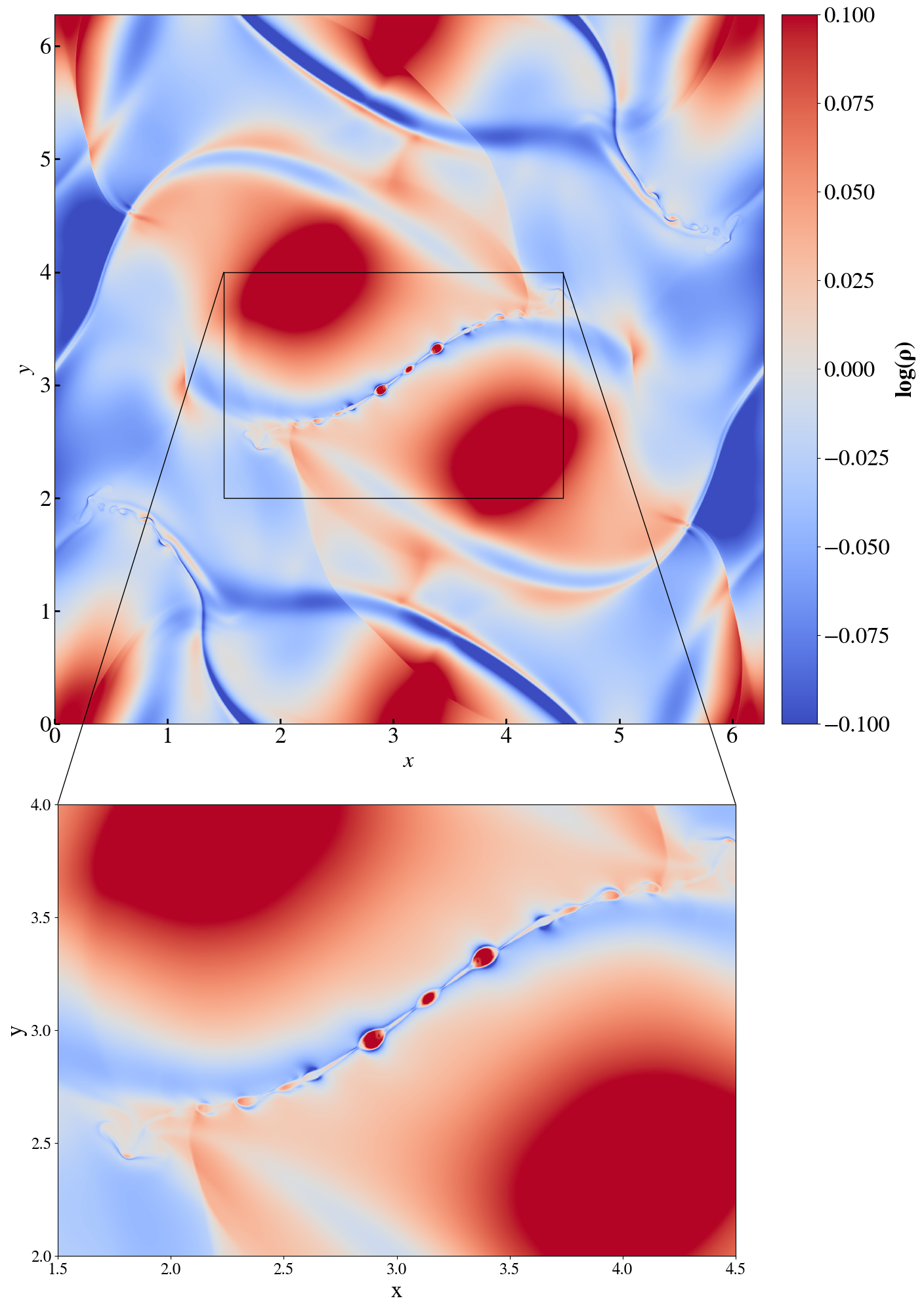}
\includegraphics[width=1\columnwidth]{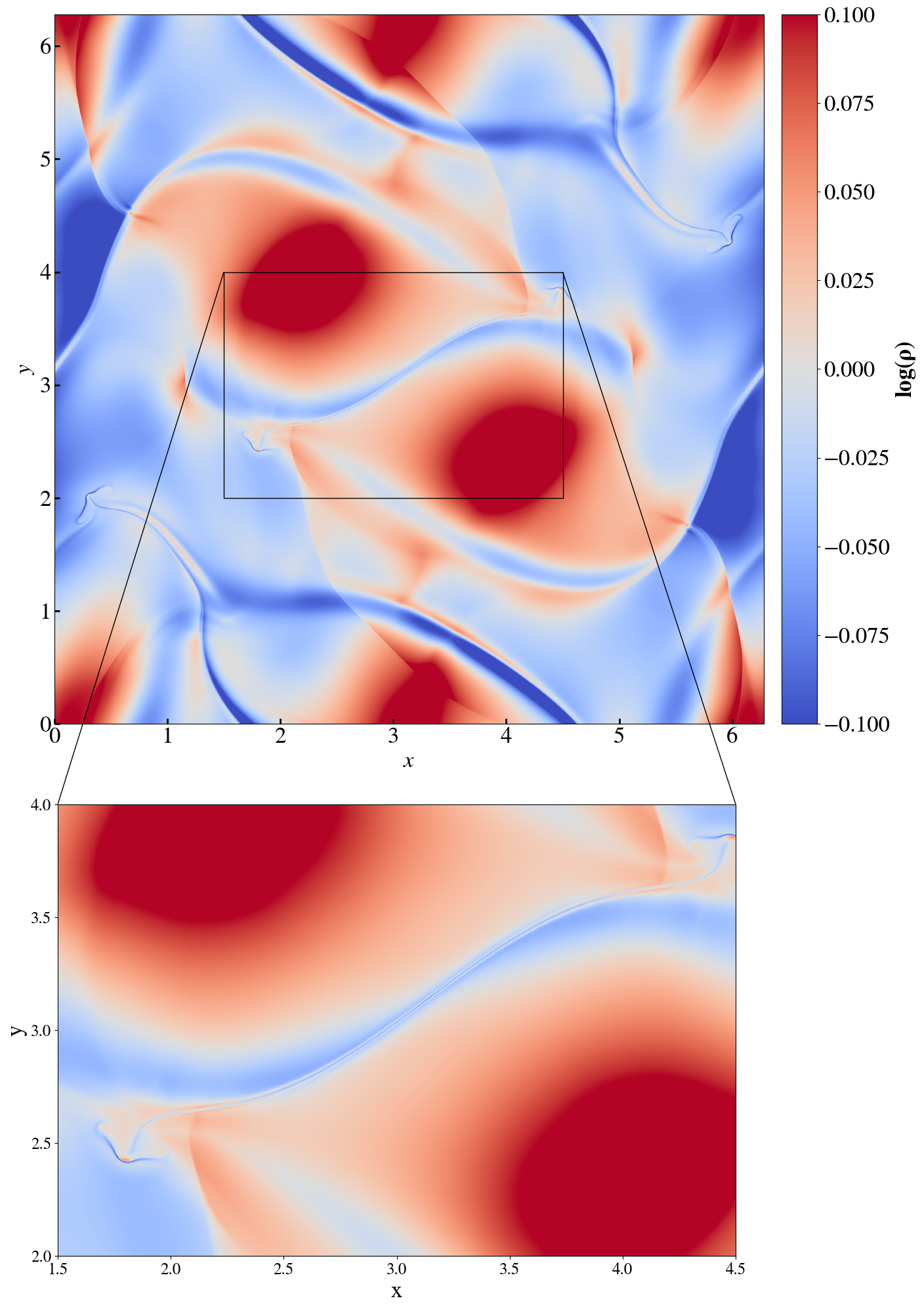}
\caption{The rest mass-density at $t=2.5t_c$ in the resolution of $4096^2$ with PLUTO in the Ideal-MHD simulations ({\it left panels}) and Res-MHD simulations with physical resistivity $\eta=10^{-4}$ ({\it right panels}). Plasmoids, zoomed-in in the bottom panels, form only in a case with sufficiently low resistivity, corresponding to a Lundquist number larger than $S\sim 10^4$.}
\label{d4096}
\end{figure*}

With implicit inclusion of the most important features of MHD turbulent flow such as energy dissipation and magnetic reconnection \citep{small-scaleorszag_tang_1979, D1989} the Orszag-Tang vortex is a comprehensive test problem for MHD codes. This problem mostly tests the code performance in simulations with MHD shocks and shock-shock interactions. 

We study the energy distribution in different setups by performing the OT test problem simulations using two astrophysical simulation codes: PLUTO \citep[ver. 4.4;][]{m07} and KORAL \citep{koral01}. The description of our simulations is mostly presented in code units. These are obtained by scaling the equations with fiducial values of certain physical quantities. All velocities are scaled with $v_0=c$, e.g.,  the statement that $v_\mathrm{A}=1$ in code units signifies that the Alfv\'en velocity is equal to the speed of light. The density is scaled with some density $\rho_0$, the pressure with $p_0$, and the electromagnetic fields with $B_0$. The exact value of $\rho_0$ is immaterial, as long as $p_0=\rho_0 v_0^2$ and $B_{0} = v_0 \sqrt{4\pi \rho_0}$.

\subsection*{Two dimensional setup}
The simulation is set up in a 2D computational box $0\leq x,y\leq2\pi$ with periodic boundary conditions and the following initial conditions for velocity and magnetic fields \citep{BR01}:
\beqa
    v = \tilde{v}(-\sin y, \sin x, 0),\label{OT01} \\
    B = \tilde{B}(-\sin y, \sin 2x, 0).\label{OT02}
\eeqa
We adopt $\tilde{v} = 0.99 v_0/\sqrt{2}$ and $\tilde{B}=B_0$. 
The initial density is uniform.

In 2D we perform the OT simulations in the range of uniform resolutions from $64^2$ to $4096^2$  in different setups (Ideal-MHD, Res-MHD, and Rel-MHD), doubling the number of grid points in each dimension to increase the resolution step by step. In 3D we run the Ideal-MHD and Rel-MHD simulations in three resolutions $128^3$, $256^3$, and $512^3$. Only with PLUTO, we run the Res-MHD simulation (in both 2D and 3D) in the resolution $512^3$. Without resistivity, both PLUTO and KORAL are used for Ideal-MHD and Rel-MHD simulations in 2D and 3D setups\footnote{The physical resistivity module is publicly available only in non-relativistic PLUTO, and this is the one we use to perform our Res-MHD simulations.}.

All simulations run to the final time $t=10\,t_c$, where $t_c$ is the light-crossing time across the typical length in the system.  In code units, $t_c=L$, and we take $t_c=1$.

\subsection*{Three dimensional setup}
In order to study the difference between 2D and 3D MHD flows and reconnection, we extend the Orszag-Tang test problem to three dimensions. 
We set up the simulation in a cubic computational box $0\leq(x, y, z)\leq 2\pi$ with periodic boundary conditions.

For the Rel-MHD simulations, the initial equations are chosen in such a way as to result in a realistic turbulent system, following the definition of a Taylor-Green vortex \citep{small-scaleorszag_tang_1979}: 
\beqa
    v = \tilde{v}(\cos z \sin y \cos z, - \sin x \cos y \cos z, 0),\label{OT03}\\
    B = \tilde{B}(-\sin y, \sin 2x, 0),\label{OT04}
\eeqa
where $\tilde{v}$ and $\tilde{B}$ are the same as in the 2D setup.

We find that such initial conditions do not result in a sufficiently turbulent outcome in non-relativistic simulations in 3D, so for Ideal-MHD and Res-MHD simulations in 3D we use different initial conditions, following \cite{Mininni_2006}: 
\beqa
    v = \tilde{v}(-\sin{y}, \sin{x}, 0),\label{OT05} \\
    B = \tilde{B} (-2\sin{2y}+\sin{z}, 2\sin{x}+\sin{z}, \sin{x}+\sin{y}),\label{OT06}
\eeqa
where $\tilde{v}=2v_0$ and $\tilde{B}=0.8 B_0$.
The initial density is uniform.

\section{Energy components in the results}
\label{resdis}
%

We study the dissipation of magnetic energy and investigate the conversion of energy by following the time evolution of the energy components: the electromagnetic energy density $U_{\rm EB}=E_{\rm B}+E_{\rm E}=\frac{1}{2}(B^2+E^2)$, 
the kinetic energy density $E_{\rm K}$, and internal energy density $U_{\rm int}$. We study all components in the laboratory frame, thus the kinetic energy and internal energy densities in the relativistic simulations Rel-MHD are computed as follows:
\beqa
    E_{\rm K} = \rho (\Gamma^2 -1)c^2,\label{eq20} \\
    U_{\rm int} = \left(\frac{\gamma}{\gamma -1} \Gamma^2  - 1\right) p.\label{eq21}
\eeqa
Here, $\gamma=4/3$ is the polytropic constant.
In the non-relativistic limit (simulations Ideal-MHD and Res-MHD) the internal energy density becomes
\beqa
    U_{\rm int} = \frac{p}{\gamma -1},\label{eq23}
\eeqa
while the kinetic energy density is
\beqa
    E_{\rm K} = \frac{1}{2} \rho \upsilon^2,\label{eq22}
\eeqa
as can be seen from Eqs.~\ref{energy-momentum2_nonrel}, \ref{en-density_nonrel}.
Another quantity that is a function of space and time is the magnetization defined as $\sigma = B^2/(\rho c^2)$.

We discuss and compare the averaged energy densities denoted by a bar and computed in 3D through
\beqa
    \overline{Q}=\frac{\iiint_V Q \,dx\,dy\,dz}{\iiint_V  \,dx\,dy\,dz},
\eeqa
where $V$ is the volume of the simulation box.
In 2D the corresponding formula is
\beqa
    \overline{Q}=\frac{\iint_S Q \,dx\,dy}{\iint_S  \,dx\,dy}.
\eeqa
 
The results in PLUTO and KORAL simulations are very similar both qualitatively and quantitatively. Unless stated otherwise, we present the PLUTO results. The KORAL results and details of their difference from the PLUTO results are discussed in detail in Section \S\ref{Scodes}.

\subsection{Ideal-MHD and Res-MHD simulations} 
\label{Sres}
In this section, we estimate the numerical dissipation in the simulations and study the effect of resistivity on the evolution of the system. In the left panel of Fig.~\ref{MHD/Res}, we plot the time evolution of the averaged squared magnetic field $\overline{B^2}$ measured in Ideal-MHD simulations for different resolutions. It is clear that at later times the value of $\overline{B^2}$ increases with an increase in the resolution. This is because in grid-based codes the flux is computed over the surface of every grid cell. In such a calculation there is some amount of computational dissipation, so-called numerical resistivity. Before we study the effect of physical resistivity in simulations, it is important to estimate the numerical dissipation at each resolution and find a reasonable minimal resolution.

We compare the results in non-resistive Ideal-MHD simulations with the Res-MHD simulations set with different physical resistivities ($\eta$ in Eq.~\ref{current_nonrel}), at each resolution\footnote{With a different setup in PLUTO, \cite{cemvt14} found that for the numerical resistivity to decrease by an order of magnitude, the number of grid cells should be quadrupled, as also follows from the estimate with the characteristic length and diffusive timescale, $\eta=L^2/t_\eta$.}. In the right panel of Fig.~\ref{MHD/Res} the results obtained with the resolution $512^2$ are shown. We compare $\overline{B^2}$ of the simulations with $\eta=0, 10^{-4}, 10^{-3}, 5\times10^{-3}$. The curves corresponding to the Ideal-MHD and Res-MHD simulations with $\eta=10^{-4}$ are almost overlapping, so at this resolution we estimate the numerical resistivity to be below $10^{-4}$ and conclude that the resolutions higher than $512^2$ are reasonably reliable for our simulations with the PLUTO code. 

The magnetic energy initially increases and then decreases, forming the hump at $2t_c$ in its plot (Fig.~\ref{MHD/Res}). This is caused by the compression of a region around a current sheet and subsequent formation of a reconnection layer (at $t\approx 2t_c$) which then dissipates the magnetic field energy.

In Fig.~\ref{d4096} we show the mass-density plots at $t=2.5t_c$ in the simulations Ideal-MHD (numerical resistivity below $10^{-4}$) and Res-MHD (physical resisistivity $\eta=10^{-4}$) for the resolution of $4096^2$. In the left panel (Ideal-MHD) we have identified plasmoids (regions of higher density and lower magnetization relative to their surroundings), these are the substructures located in the central region of the simulation box. In the right panel (Res-MHD) the chain of plasmoids does not appear. Similarly, we see no such chain in the simulations with a resistivity larger than $10^{-4}$. The resistivity of $10^{-4}$ corresponds to the Lundquist number $S=L v_A/\eta \,\approx\,10^4$, with the typical length scale of the system $L\,\approx\,1$ and Alfv\'en velocity $v_\mathrm{A} \,\approx\, 1$. This result matches theoretical studies which confirm that the current sheet is plasmoid unstable\footnote{Plasmoid unstable current sheet involves a dynamic process where plasmoids merge and split within a sheet-like structure of magnetized plasma.} at $S > 10^4$ \citep{Loureiro_2007, BR01}. We also confirm that with a smaller physical resistivity ($\eta < 10^{-5}$, $S > 10^5$) some substructures are resolved in the Res-MHD simulations.

\begin{figure*}
    \centering
    \includegraphics[width=2\columnwidth]{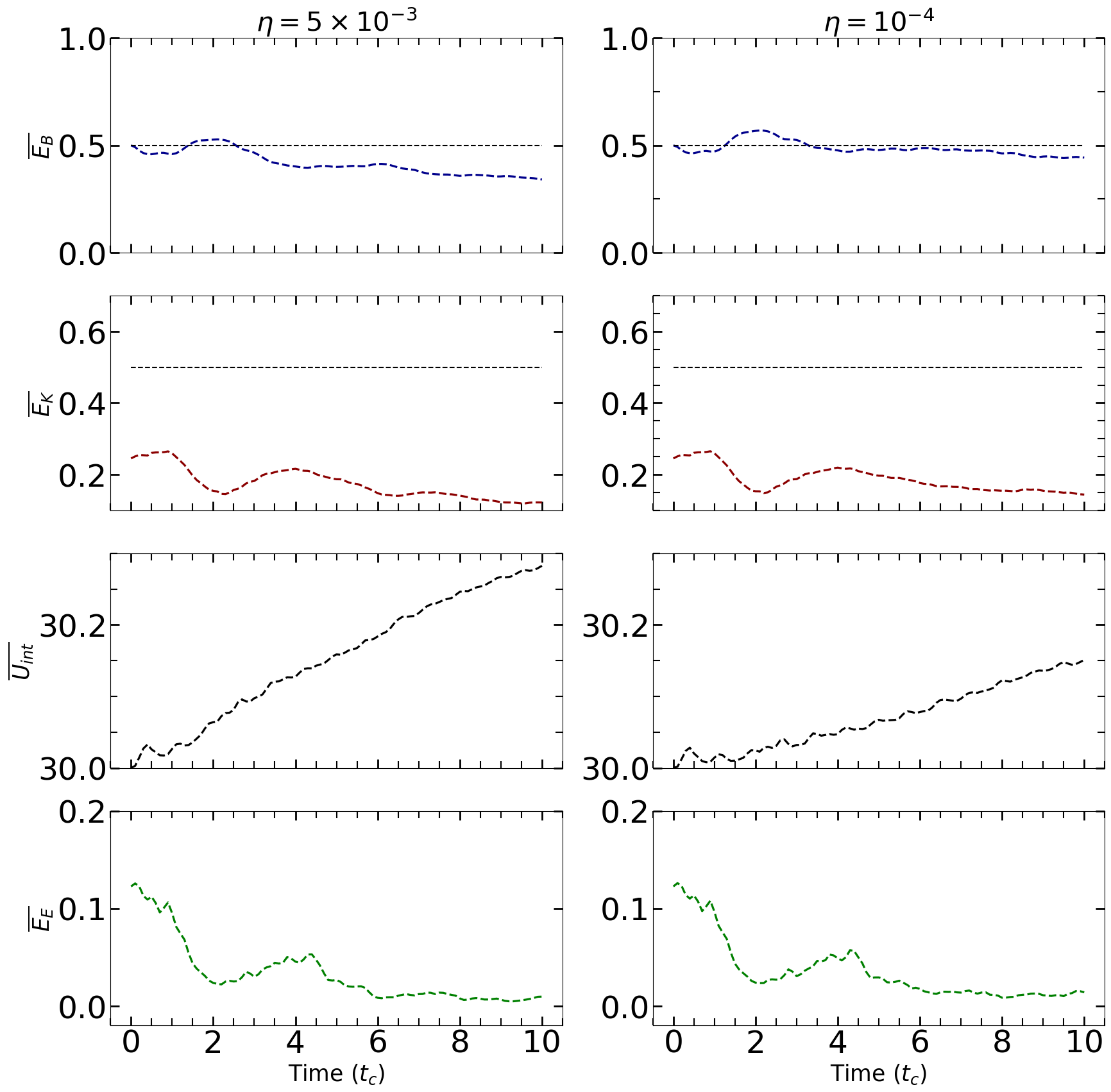}
    \caption{Energy distribution in Res-MHD simulations with physical resistivities $\eta = 5\times 10^{-3}$ ({\it left panels}) and $\eta = 10^{-4}$ ({\it right panels}) at the resolution of $512^2$ grid cells. The horizontal black dashed lines in the $\overline{E_B}$ and $\overline{E_K}$ panels indicate the initial value (0.5) of the magnetic energy $\overline{E_B}$. See the detailed discussion in \S\ref{Sres}. The dissipated magnetic energy heats up the plasma.}
    \label{residual}
\end{figure*}

We compare the different terms in energy distribution (magnetic energy $\overline{E_B}$, kinetic energy $\overline{E_K}$, internal energy $\overline{U_{ \rm int}}$, and electric energy $\overline{E_E}$, respectively) in the Res-MHD simulations with  $\eta = 5\times 10^{-3}$ and $\eta=10^{-4}$ (Fig.~\ref{residual}). The first row of this figure shows magnetic energy where the horizontal dashed line, located at $\overline{E_B}=0.5$, shows the initial value of magnetic energy. We see that with decreasing physical resistivity (from the left panel to the right panel) the rate of magnetic energy decrease becomes smaller. 
The dissipated magnetic energy converts to the internal energy and heats up plasma as shown in the third row of this plot. We will discuss the energy components in Rel-MHD and Ideal-MHD simulations in the next section.
%
\begin{figure*} 
\centering
\includegraphics[width=2\columnwidth]{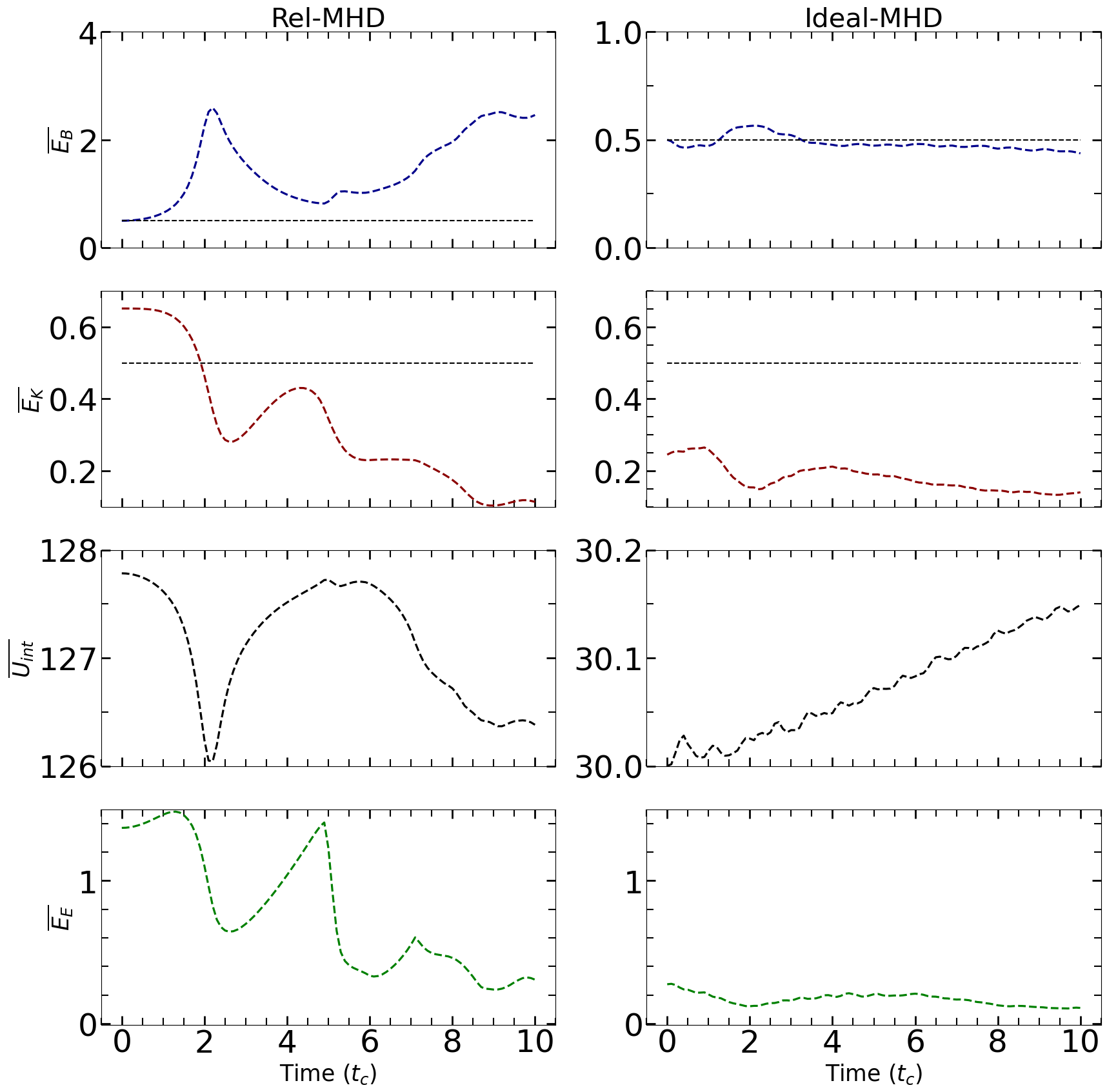}
\caption{Energy distribution in Rel-MHD and Ideal-MHD PLUTO simulations at the resolution of $512^2$ grid cells are shown in the ({\it left}) and ({\it right}) panels, respectively. The horizontal black dashed lines in the panels with $\overline{E_B}$ and $\overline{E_K}$ indicate the initial value of the magnetic energy $\overline{E_B}=0.5$. See the detailed discussion in \S\ref{Srel}.}
\label{rnr}
\end{figure*}

%
\begin{figure}
\includegraphics[width=1\columnwidth]{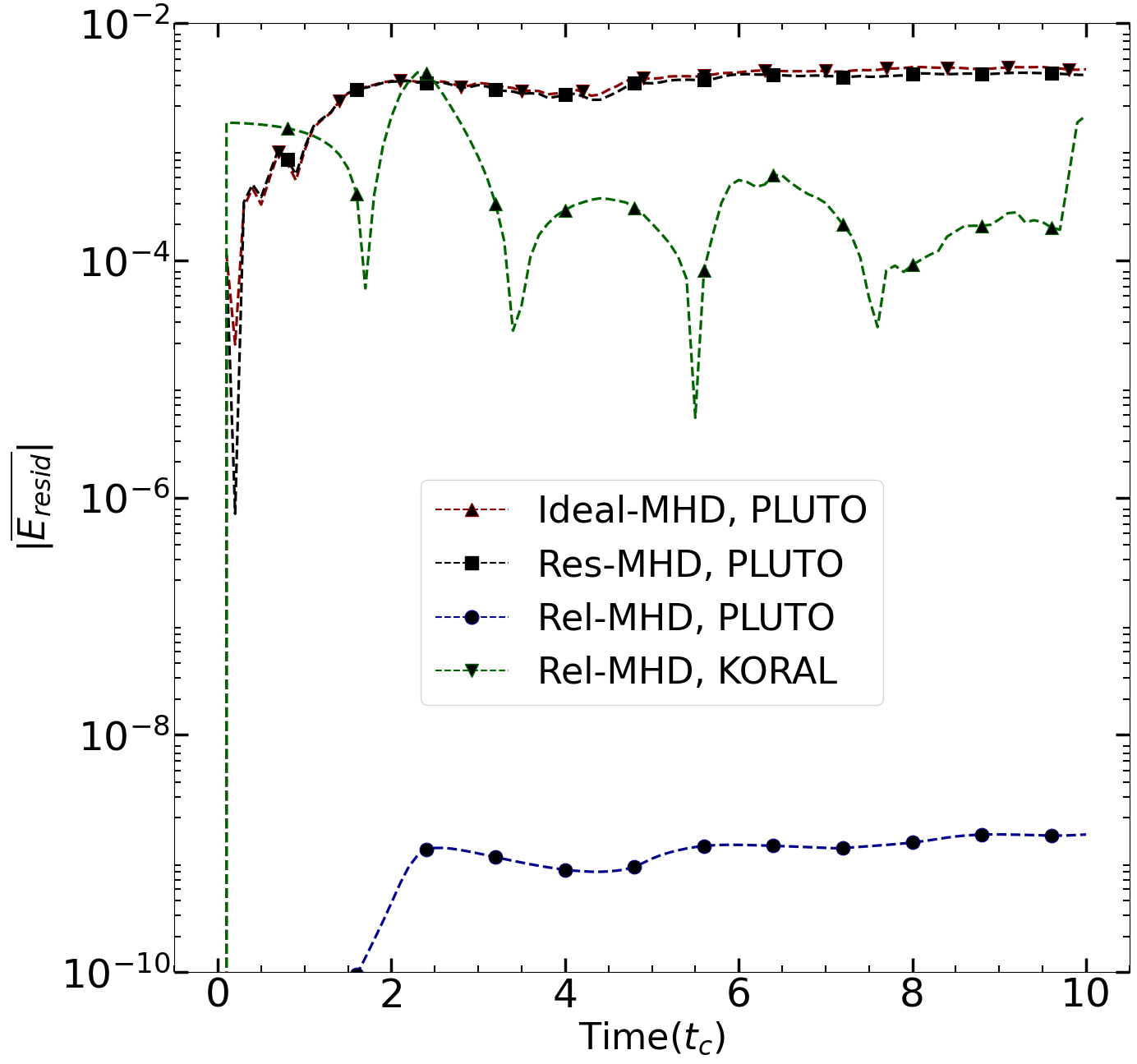}
\caption{Residuals of total energy $\overline{|E_{\rm resid}|}=|E_{\rm tot}(t)-E_{\rm tot}(0)|/E_{\rm tot}(0)$ in the Ideal-MHD and Res-MHD ($\eta = 5\times10^{-3}$) simulations with PLUTO (top two, nearly coinciding curves) and Rel-MHD simulations with PLUTO (bottom curve) and KORAL, with the resolution $512^2$ grid cells. The conservation of energy is significantly more accurate in the Rel-MHD simulation, particularly in PLUTO.}
\label{totrnr}
\end{figure}
%
\begin{figure}
\includegraphics[width=1\columnwidth]{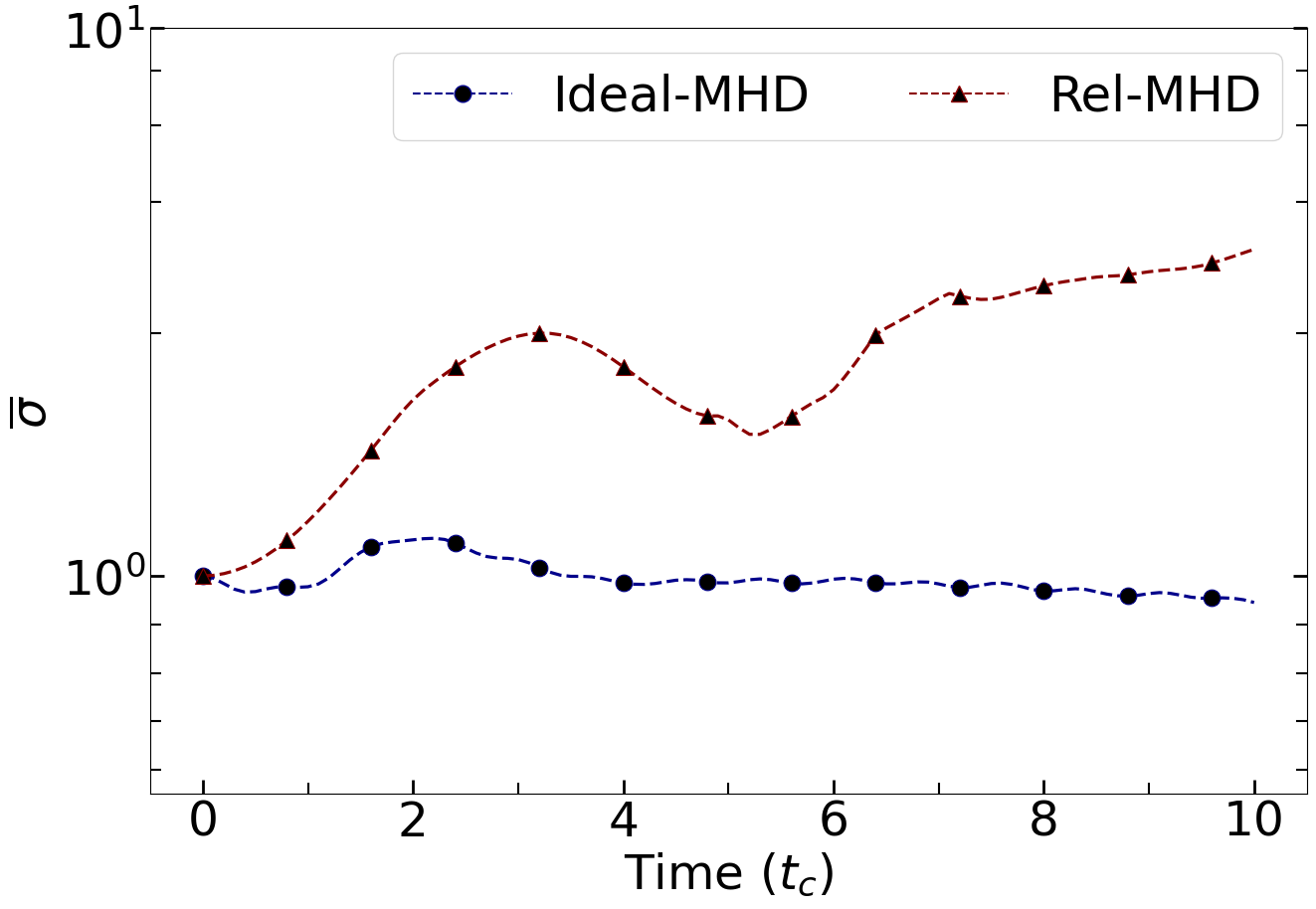}
\caption{Magnetization ($\sigma = B^2/\rho$ in code units) in Ideal-MHD and Rel-MHD simulations with PLUTO at the resolution $512^2$ grid cells.}
\label{sigmarnr}
\end{figure}
%
\subsection{Ideal-MHD and Rel-MHD simulations}
\label{Srel}
We compare the results of non-relativistic (Ideal-MHD) and relativistic (Rel-MHD) non-resistive MHD simulations in the PLUTO code. The different terms in energy distribution 
(magnetic energy $\overline{E_B}$, kinetic energy $\overline{E_K}$, internal energy $\overline{U_{ \rm  int}}$, and electric energy $\overline{E_E}$, respectively) are shown in Fig.~\ref{rnr}. Panels in the left column show the results for Rel-MHD and in the right column for Ideal-MHD. 

The magnetic energy evolution, shown in the first row of panels in Fig.~\ref{rnr}, indicates that in simulation Rel-MHD the magnetic energy increases five-fold from the initial value of $0.5$ (shown by the black dashed line in both left and right top panels). In the non-relativistic simulation Ideal-MHD there is only a minor initial increase of the magnetic energy followed by a slow decay.

The kinetic energy evolution is presented in the second row of Fig.~\ref{rnr}, where a black dashed line is also drawn for reference at the value of 0.5. The kinetic energies were computed using Eqs. \ref{eq20} and \ref{eq22} for the Rel-MHD and Ideal-MHD simulations, respectively. In Rel-MHD the effect of the Lorentz factor on the kinetic energy leads to an initial value of approximately 0.62, which is higher than the magnetic energy. In contrast, for the Ideal-MHD simulations, the initial value of the kinetic energy is approximately 0.25, half the value of the magnetic energy. Initially, in Rel-MHD the kinetic energy amplifies the magnetic field, while in the nonrelativistic Ideal-MHD case the low value of $\overline{E_{\mathrm K}}$ is not enough to amplify the magnetic energy. Thus, in Rel-MHD the effect of kinetic energy on the magnetic energy evolution in the second half of the simulation is significant, causing a secondary increase of $\overline{E_{\mathrm B}}$. In the Ideal-MHD no such effect is observed.

In the third row of panels in Fig.~\ref{rnr}, we show the internal energy as computed from Eqs. \ref{eq21} and \ref{eq23}. Comparison with the first row of panels shows the conversion between magnetic energy and internal energy. 

In Rel-MHD, after $t\simeq 5t_c$, the large amount of the internal and kinetic energy amplifies the magnetic field. This is visible as the second increase (``hump'') in the $\overline{E_B}$ curve.
Such an outcome in the Rel-MHD simulation offers an explanation for the energy reservoir in magnetized systems like relativistic jets in active galactic nuclei, accretion discs of black holes, and magnetized neutron stars in high-energy astrophysics. 
In the nonrelativistic Ideal-MHD case, shown in the right panel, the released magnetic energy converts to internal energy and heats up the plasma. In contrast with the relativistic case, the amount of energy in the system is not enough to re-amplify the magnetic field. 

The final row of panels in Fig.~\ref{rnr} displays the electric energy, which exhibits a significant evolution in the Rel-MHD simulation. The electric field is a function of magnetic field and velocity (Eq.~\ref{idealfield}), Consequently when the magnetic field is increased around $\sim 2t_c$, the electric energy $\overline{E_E}$ also increases. Furthermore, as the system evolves, there is another subsequent increase in $\overline{E_E}$, coinciding with an increase in kinetic energy after $4t_c$. 

The sum of all energy components in each of the simulations is conserved over time, as shown in Fig.~\ref{totrnr}. The residuals in the total energy, $\overline{|E_{\rm resid}(t)|}=|E_{\rm tot}(t)-E_{\rm tot}(0)|/E_{\rm tot}(0)$, are displayed for the simulations Ideal-MHD, Res-MHD (with $\eta = 5\times 10^{-3}$), and Rel-MHD. Here, $E_{\rm tot}$ is the sum of the magnetic, electric, kinetic, and internal energies displayed in Fig.~\ref{totrnr}. The residuals in the Ideal-MHD and Res-MHD in PLUTO simulation are $\approx 3\times 10^{-3}$, in KORAL Rel-MHD $\sim 10^{-4}$, whereas, in the Rel-MHD case in PLUTO they are $\approx 10^{-9}$. This indicates that the numerical dissipation in the relativistic simulation is significantly lower than in the non-relativistic simulations. Also, the results indicate that Rel-MHD simulation in PLUTO is less dissipative than in KORAL. 

Space averaged magnetization in both simulations, Rel-MHD and Ideal-MHD, with the fixed resolution of $512^2$ grid cells, is shown as a function of time in Fig. \ref{sigmarnr}. This shows once again how the relativistic system is strongly magnetized and the magnetization increases by the end of the simulation time, while in a non-relativistic simulation, the magnetization does not evolve significantly.
\subsection{3D simulations}
%
\begin{figure*}
\centering
\includegraphics[width=1\columnwidth]{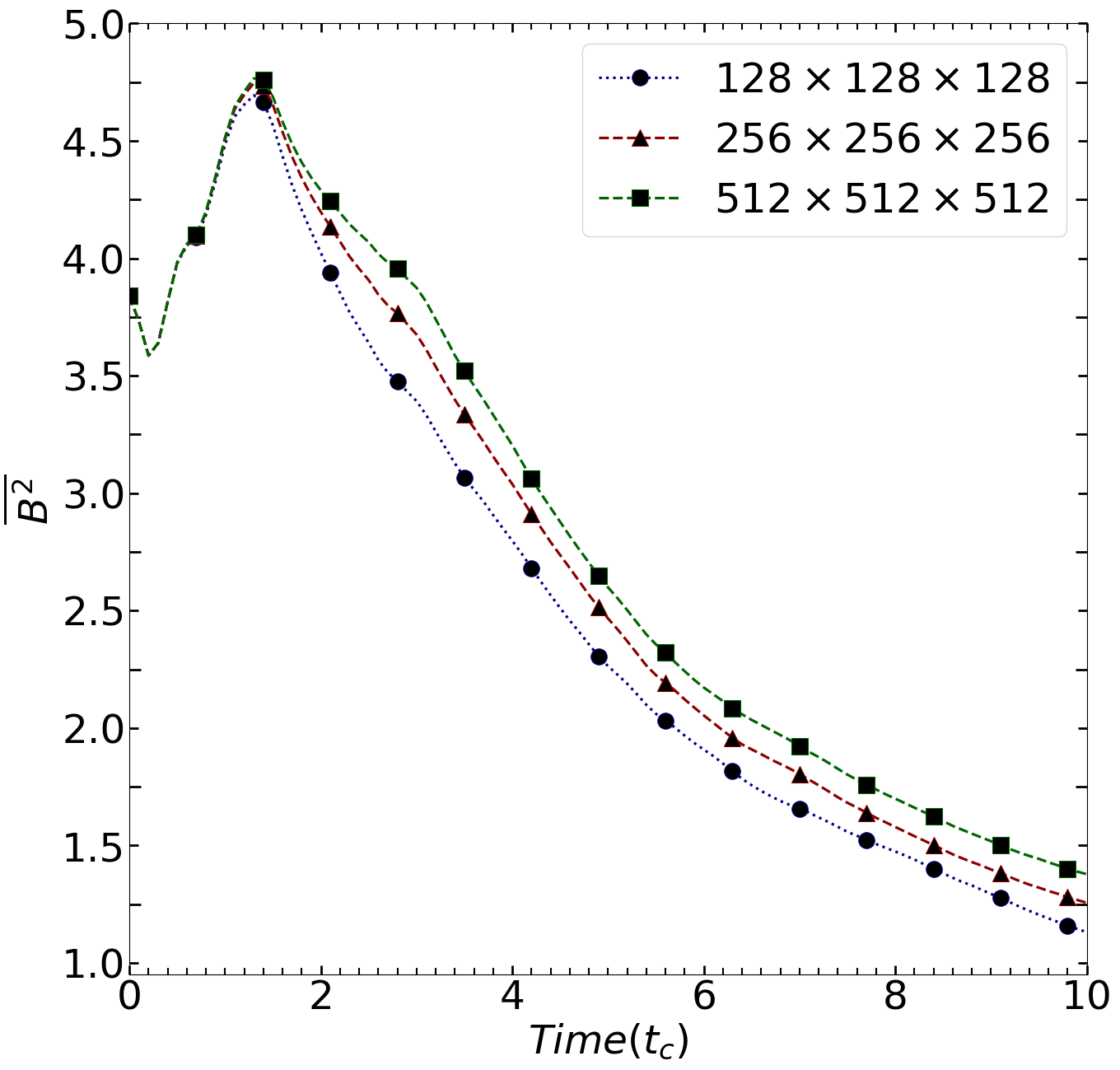}
\includegraphics[width=1\columnwidth]{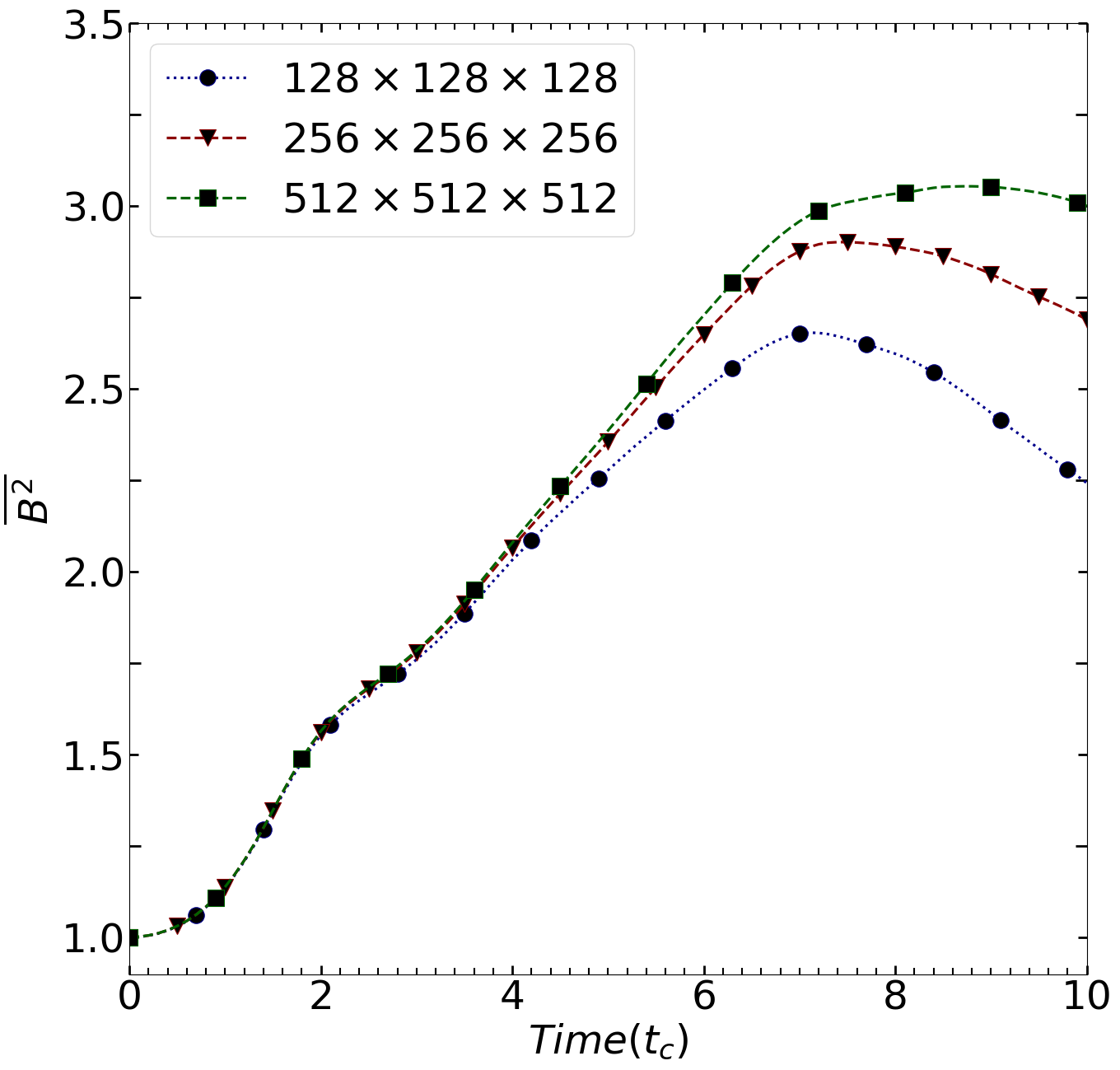}
\caption{The time evolution of $\overline{B^2}$ in 3D simulations with  PLUTO for the simulations Ideal-MHD ({\it left panel}) and Rel-MHD ({\it right panel}).}
\label{3drnr}
\end{figure*}
%
\begin{figure*}
\centering
\includegraphics[width=1\columnwidth]{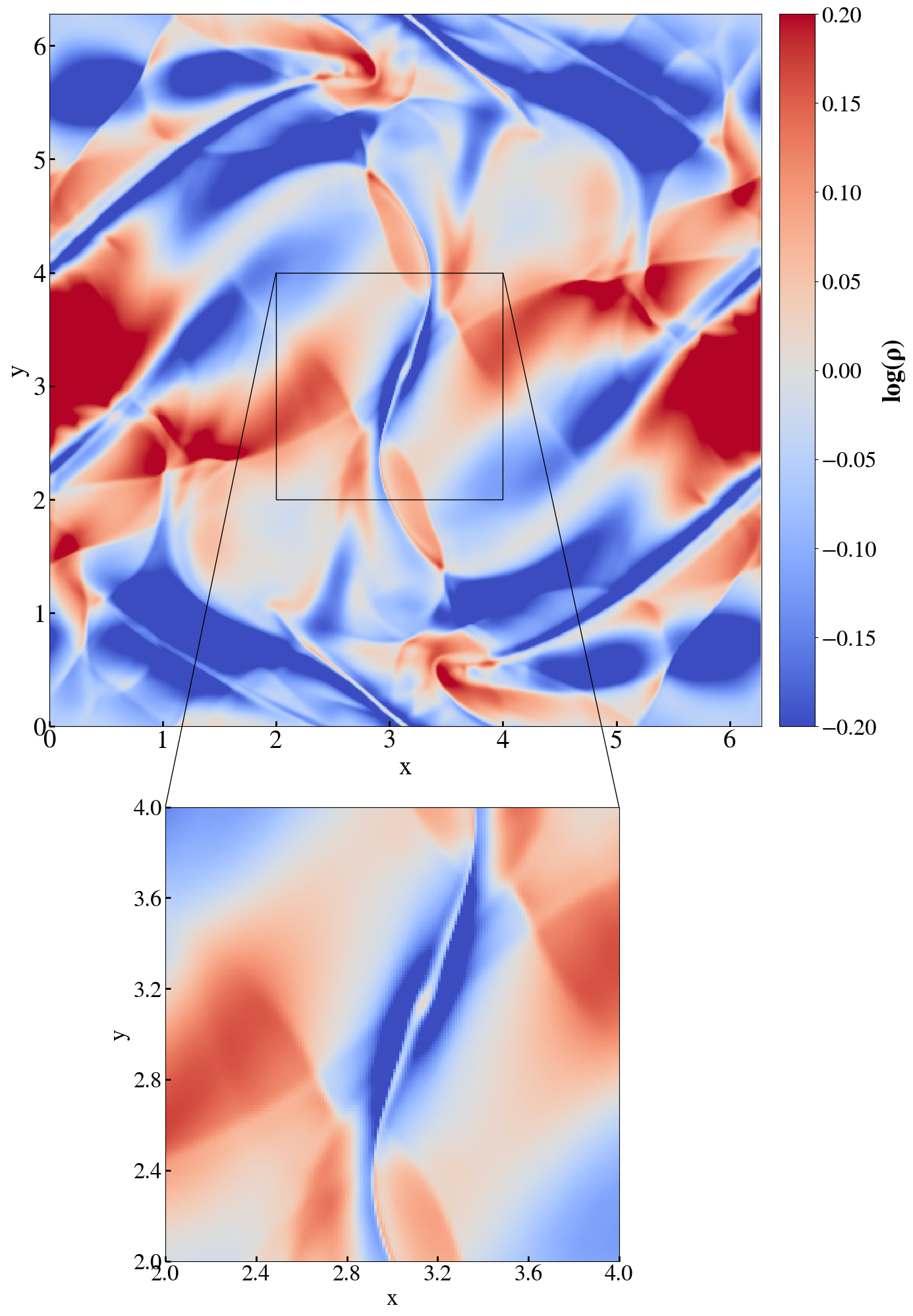}
\includegraphics[width=1\columnwidth]{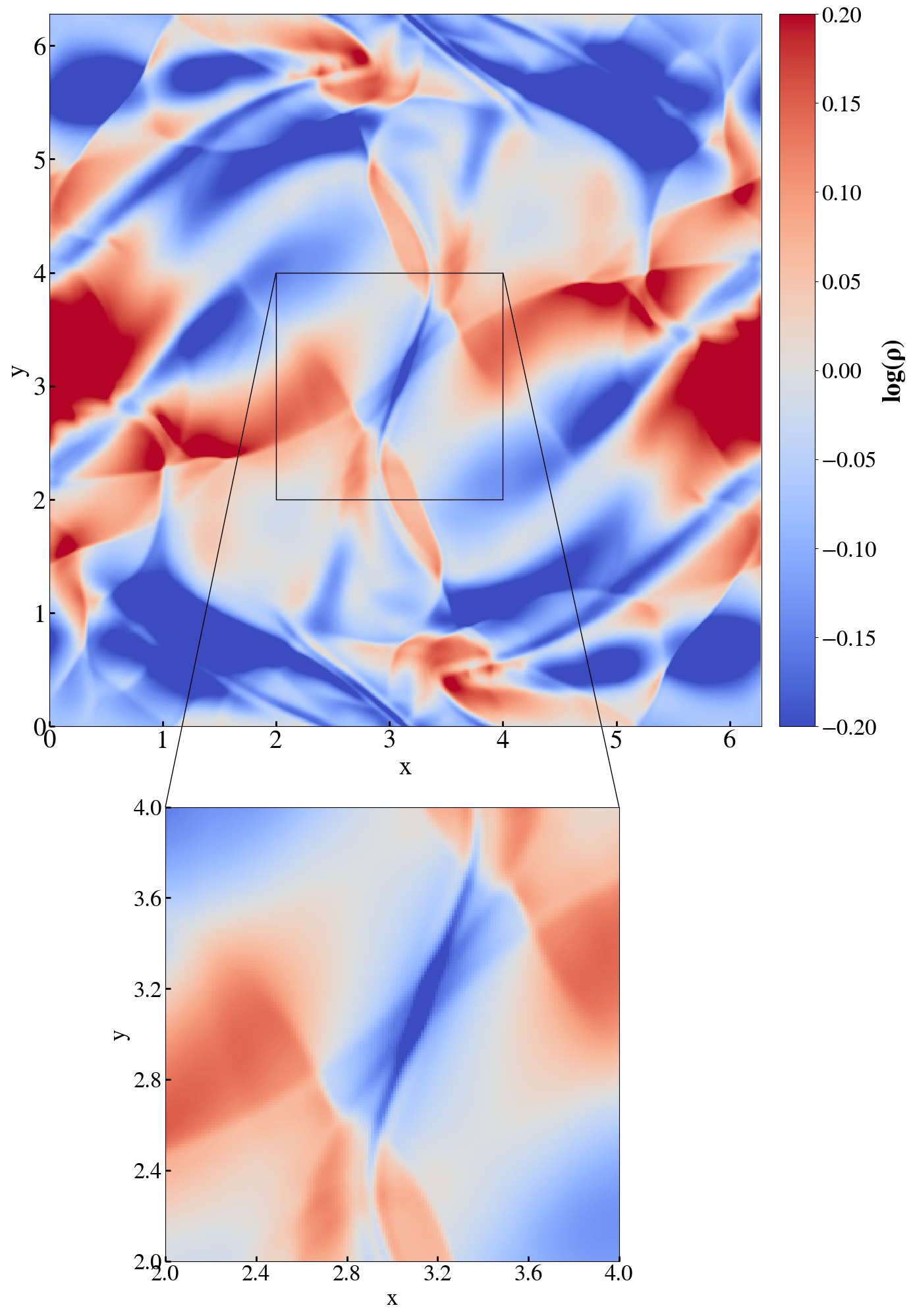}
\caption{The slice in $z=\pi/2$ in the simulation box of the rest-mass density $ \rho$ for a vortex at $t=1.5t_c$ at the resolution of $512^3$ in PLUTO. {\it Left panel}: Ideal-MHD. {\it Right panel}: Res-MHD with $\eta = 10^{-3}$. The zoomed-in panels show the current layer in the middle of the simulation boxes. Plasmoids form only in the cases with sufficiently low resistivity, corresponding to a Lundquist number larger than $10^4$ ($\eta \lesssim 10^{-4}$).}
\label{Res3d_density}
\end{figure*}
%
\begin{figure}
\includegraphics[width=1\columnwidth]{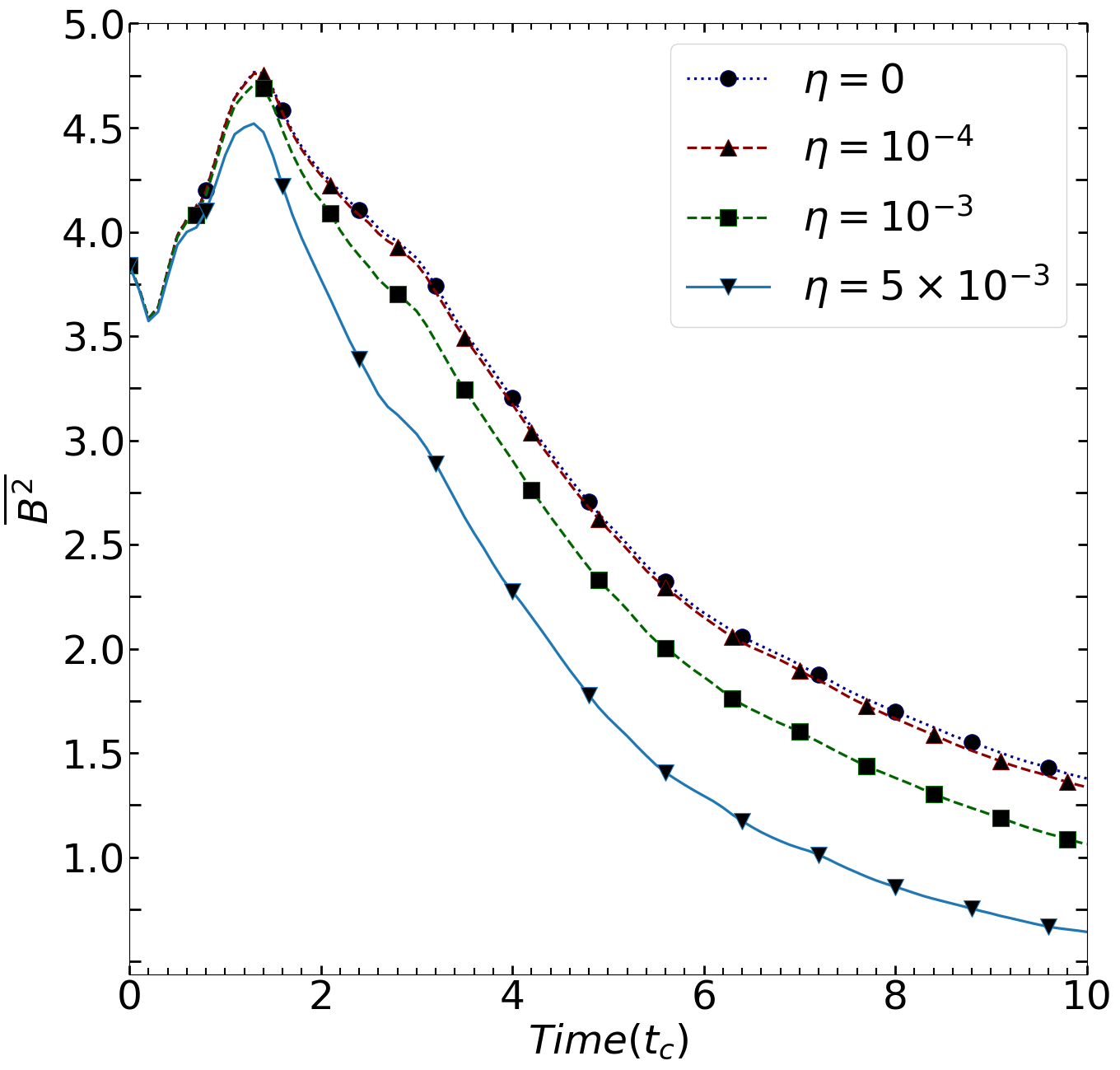}
\caption{The time evolution of $\overline{B^2}$ in 3D Res-MHD simulations with PLUTO, at the resolution $512^3$ with different physical resistivities.}
\label{Res3D}
\end{figure}
%

We perform the OT test problem simulations in three dimensions in PLUTO and KORAL with the initial conditions of Eqs. \ref{OT03}, \ref{OT04} in Rel-MHD simulations and with the initial conditions of Eqs. \ref{OT05}, \ref{OT06} in Ideal-MHD and Res-MHD simulations. 

The time evolution of $\overline{B^2}$ in the Ideal-MHD simulations is shown in the left panel of Fig. \ref{3drnr}. We expect the current sheet to be resolved at time $t\simeq 1.5 t_c$, because of the increase in magnetic energy discussed in the previous section. We search for the reconnection layers and plasmoids in different slices of the simulation domain at this simulation time. An example of a resulting rest-mass density plot is shown in Fig.~\ref{Res3d_density}, which is a slice at $z=\pi/2$. The plasmoid (in the left panel) is shown at the center of the simulation box, which is zoomed-in at the bottom panel. 

We estimate the numerical resistivity at each resolution in Ideal-MHD simulations in 3D by comparing with Res-MHD simulations for different values of $\eta$. The plot of $\overline{B^2}$ with different physical resistivities $\eta=0, 10^{-4}, 10^{-3}, 5\times10^{-3}$, in the resolution of $512^3$ grid cells is shown in Fig. \ref{Res3D} (the method is discussed in \S\ref{Sres}). It is shown that the curves corresponding to $\eta = 10^{-4}$ resistive simulations and the non-resistive Ideal-MHD cases are convergent, so the numerical resistivity in Ideal-MHD simulations with PLUTO at the given resolution is estimated to be $\lesssim 10^{-4}$. We expect that at this resolution the current sheets are well resolved.

The rest-mass density plots in the Ideal-MHD simulations (left panel) and resistive Res-MHD simulations with $\eta = 10^{-3}$ (right panel) with the resolution of $512^3$ are shown in Fig. \ref{Res3d_density}. The zoomed-in frames in the bottom panels show the substructure at the center of each simulation box. From the configuration of the magnetic field which is not shown in this figure, we found that there is a thick current sheet containing a plasmoid in the Ideal-MHD simulation, which is not resolved in the Res-MHD simulation. 
%
\begin{figure*}
\centering
\includegraphics[width=1\columnwidth]{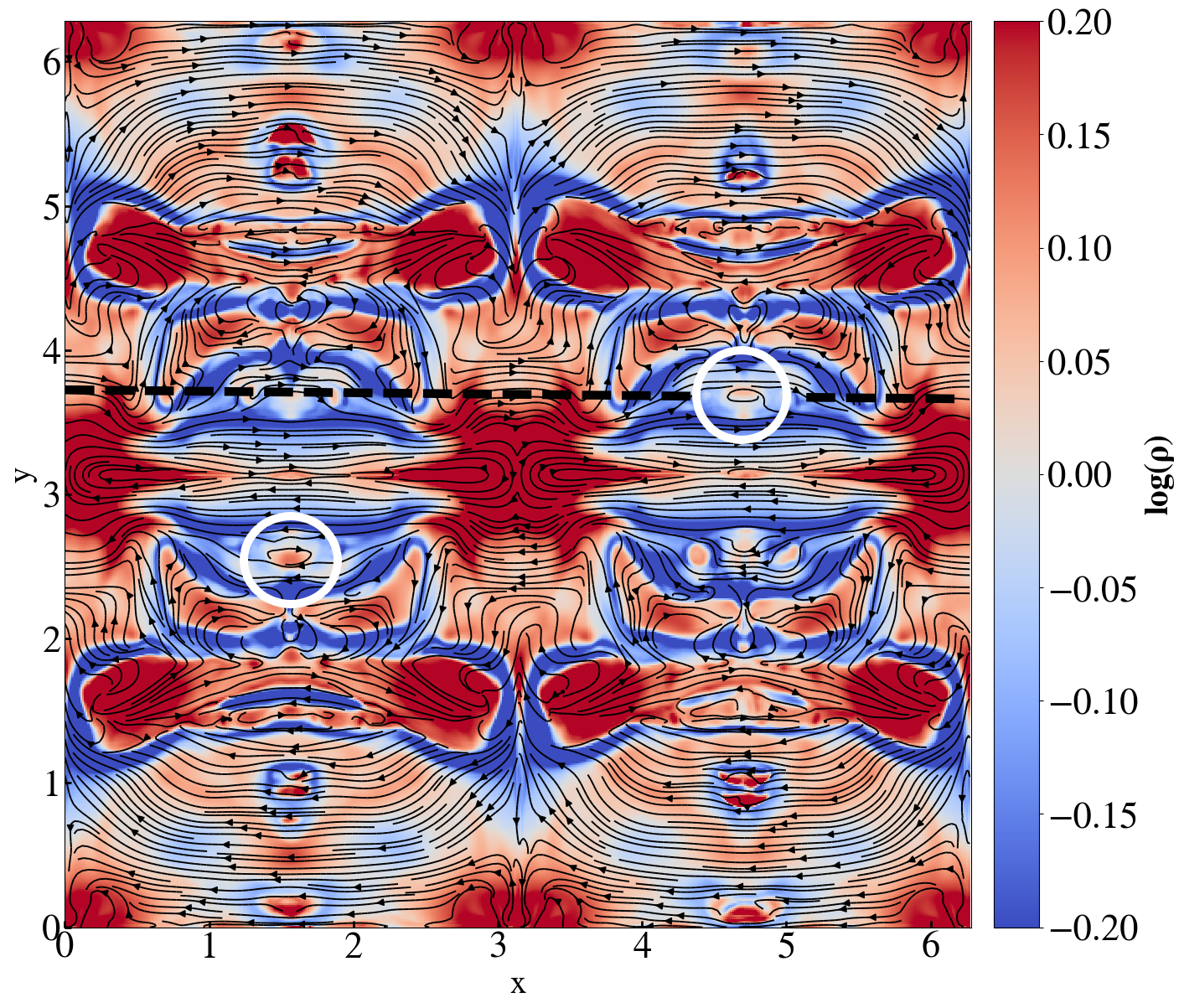}
\includegraphics[width=0.95\columnwidth]{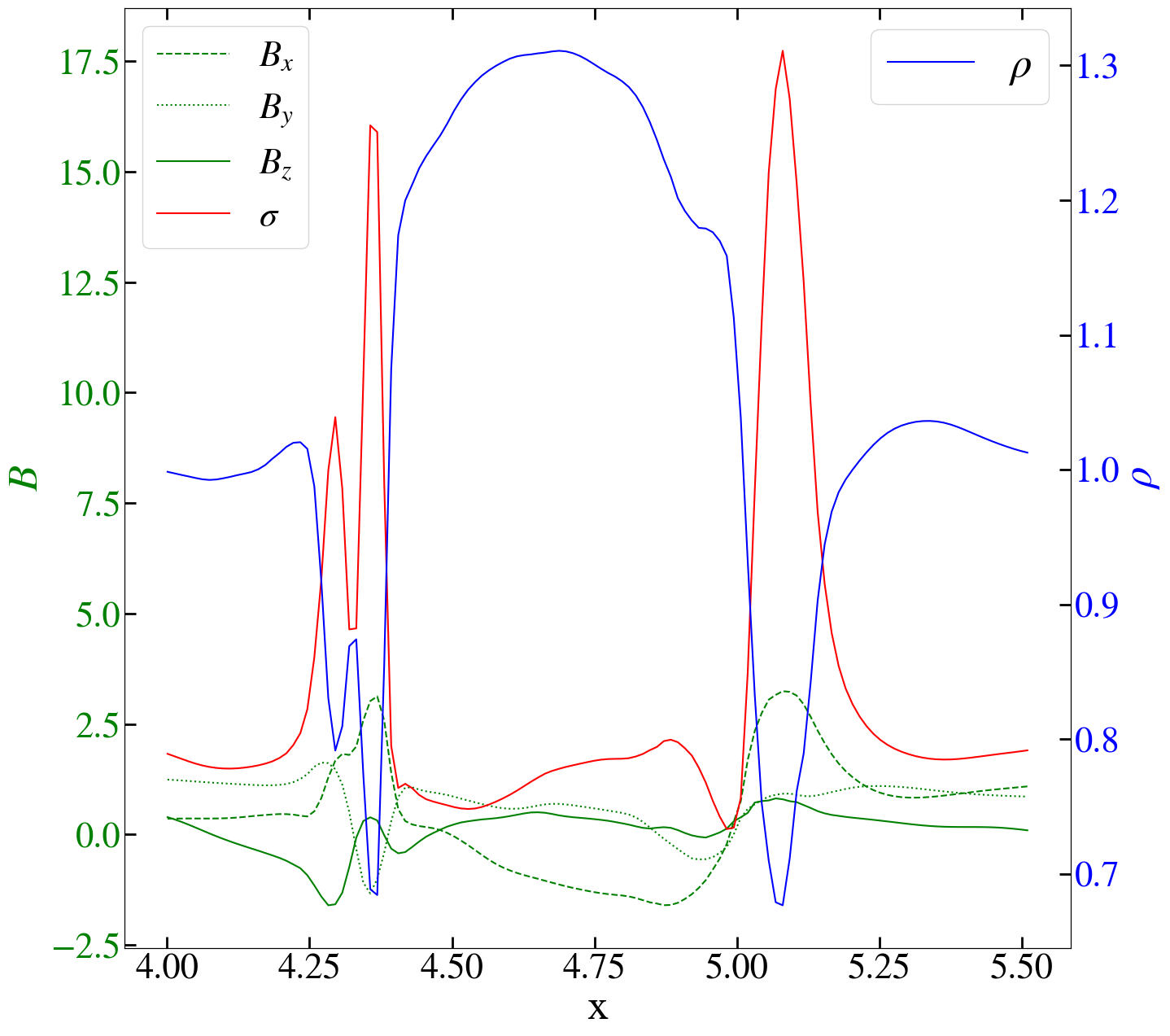}
\caption{{\it Left panel}: a slice of the rest-mass density at $z=\pi/2$ in the  Rel-MHD simulation in 3D at a resolution $512^3$ with PLUTO. The streamlines indicate the magnetic field lines and the white circles show plasmoids. {\it Right panel}: the magnetic field components, magnetization, and density profile along the black dashed line at $y=3.68$, shown in the left panel.}
\label{3Dzslice2}
\end{figure*}
%

The right panel in Fig. \ref{3drnr} shows the time evolution of $\overline{B^2}$ in the Rel-MHD simulation. It shows that $\overline{B^2}$ increases to the time $t \simeq 7t_c$. At the low resolutions, the magnetic energy drops after this time, but at the high resolution $512^3$, the peak is flattened. We found that at the smaller resolutions, due to the high numerical dissipation, the current sheets are compressed and plasmoids are not resolved. At the high resolution $512^3$, we can see the plasmoid unstable current sheets at different slices in the simulated cubic computational domain.

We show the slice in the rest-mass density at $z=\pi/2$ in the Rel-MHD simulation with the resolution $512^3$ in Fig.~\ref{3Dzslice2}, with a few magnetic islands in the simulation box\footnote{Magnetic islands are 2D slices through the structures that are plasmoids in 3D.}. We check the profile of magnetic field components and magnetization in that region. For instance, we take a closer look at one plasmoid located at $(x,y)=(4.7, 3.68)$. In the right panel, we show the profile of magnetic field components, magnetization, and mass density along the dashed line at $y=3.68$ with $x\in[4,5.6]$. The mass density $\rho$ reaches a local maximum at the position of the plasmoid, while the parallel magnetic field component $B_x$, and magnetization $\sigma$ have a minimum local value. Such a profile confirms that there is a plasmoid at this point \citep{Nathanail20, Cemeljic_2022}. In the same Rel-MHD simulation we made another slice, shown in Fig.~\ref{3Dyslice}, through the same simulation box in the $xz$ plane at $y=3.68$ (where the black dashed line is in Fig.~\ref{3Dzslice2}). In the top panel we show the reconnection layer and plasmoids. The zoomed plots show the magnetization of the selected reconnection layer. In the next section, we estimate the reconnection rate at this chosen layer. 

Using the same method (just described for the 3D Rel-MHD simulation in the last paragraph), we choose the layer shown in Fig.~\ref{3Dzslice} in the 3D Ideal-MHD run.

\section{Reconnection rate} 
\label{vr}
\begin{figure}
\centering
\includegraphics[width=1\columnwidth]{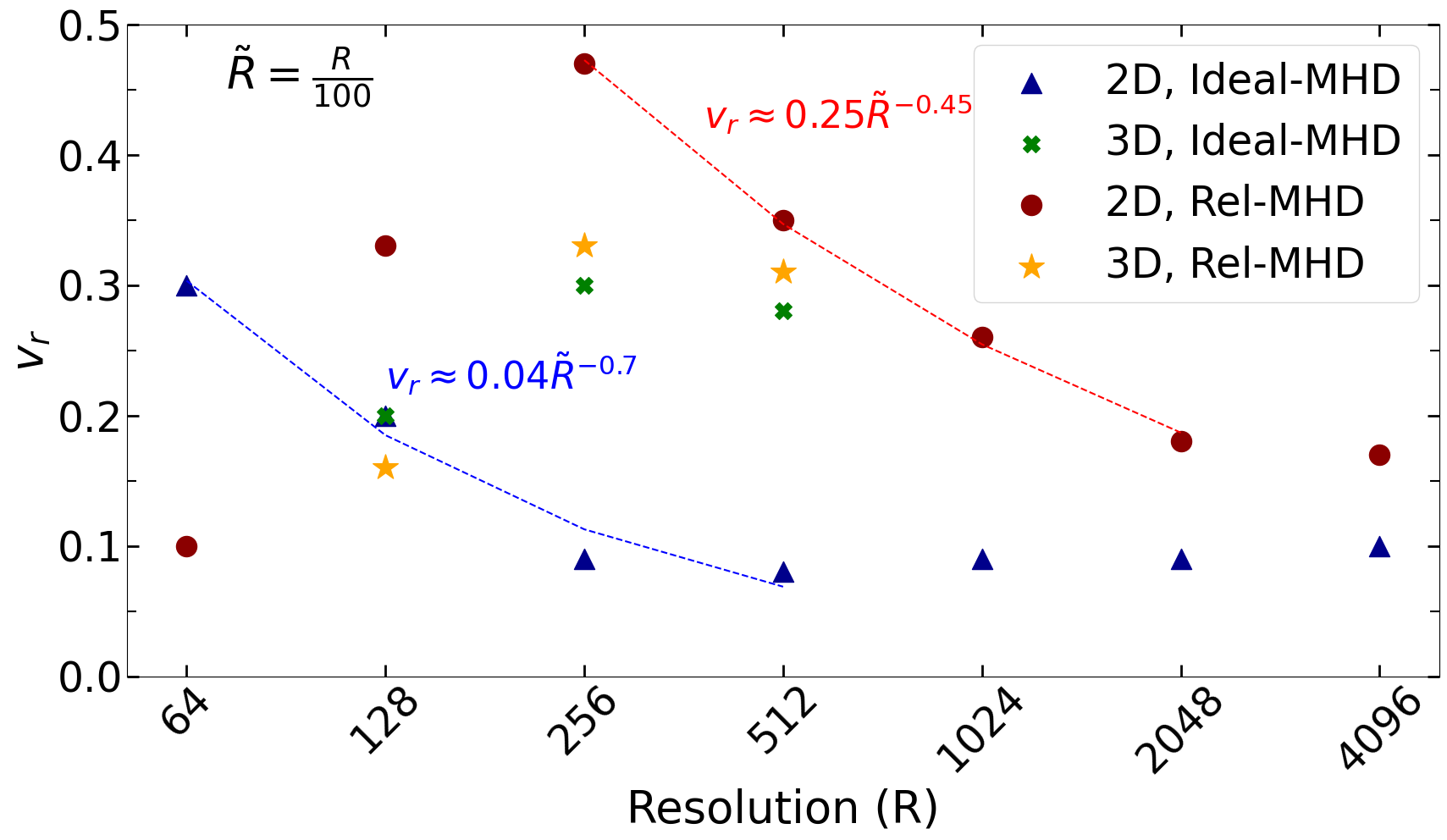}
\caption{The reconnection rate as a function of resolution in the simulations Ideal-MHD and Rel-MHD with KORAL.}
\label{vr_R}
\end{figure}
%
\begin{figure}
\centering
\includegraphics[width=1\columnwidth]{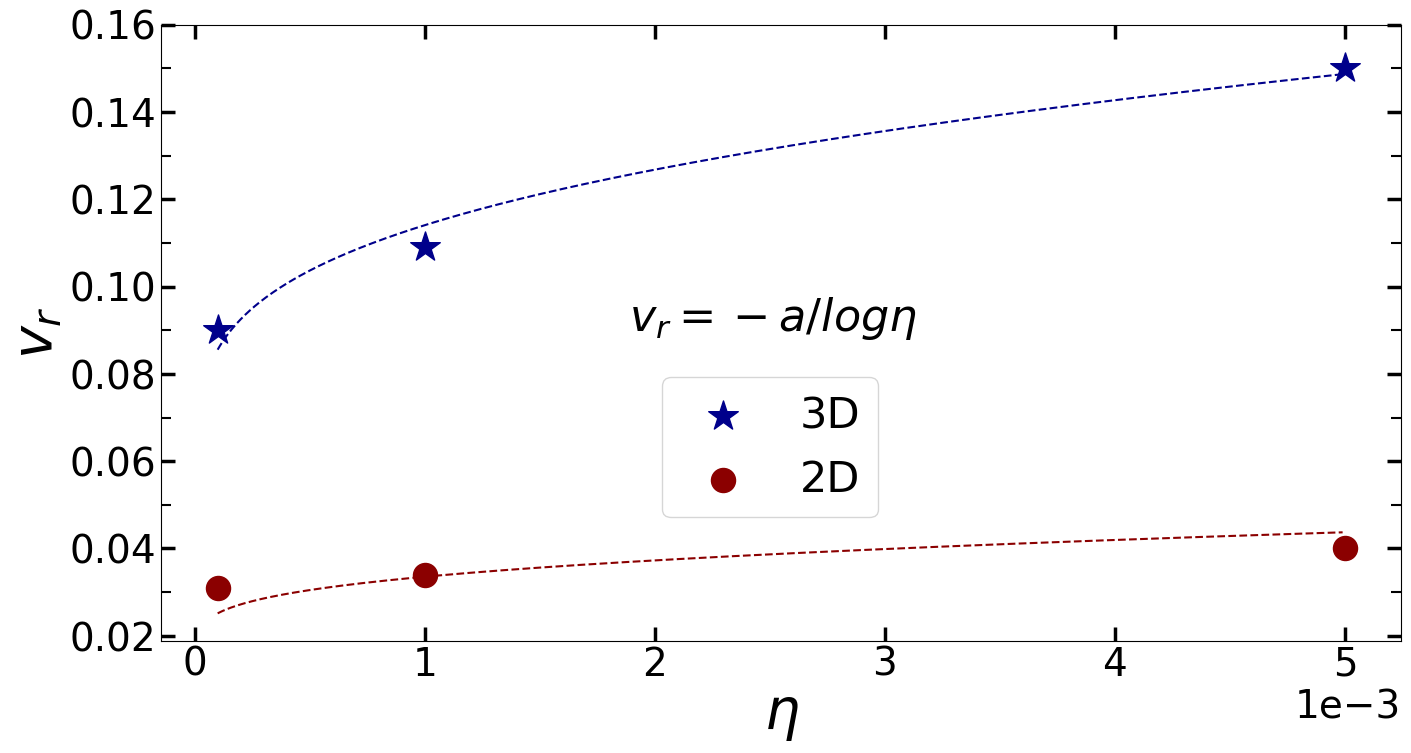}
\caption{The reconnection rate as a function of resistivity for resistivities $\geq 10^{-4}Sweet1958$ in 2D (red circles) and 3D (blue stars) Res-MHD simulations with PLUTO. The change is cosistent with $1/\log S$ dependence, $S$ being the Lundquist number.}
\label{vr_eta}
\end{figure}
%
Magnetic reconnection might occur spontaneously due to the internal MHD instability in a resistive model \citep{Sweet1958, Petscheck1964} or in the ideal MHD as a kink mode \citep{Baty_2000}. In a turbulent system, the external perturbation can cause magnetic reconnection in a so-called forced reconnection, where the plasma is in a state of chaotic and unpredictable motion. The magnetic field lines can become distorted and twisted, leading to reconnection \citep{Vekstein1998}.

Turbulent systems can be found in various environments, such as in the solar wind, in the interstellar medium, and in the accretion disks of black holes and neutron stars. In these environments, magnetic reconnection can lead to a variety of phenomena, such as the acceleration of particles to high energies, the formation of jets and flares, and the heating of the plasma. An external perturbation in turbulent plasma can accelerate the formation of the X-point, causing a  reconnection one order of magnitude faster than spontaneous reconnection. Such a reconnection process is complex and still not well understood, and is an active area of research in astrophysics and plasma physics. There are analytical and numerical studies on forced magnetic reconnection including perturbation in the isolated current sheet \citep{Vekstein1998, Potter2019}, and a study searching for the observational signatures of simulated forced reconnection in solar flares \citep{Srivastava_2019}. 

The OT is a vortex problem, for which turbulence develops during evolution. It is shown in the rest-mass density plots (Figs. \ref{dmhd} and \ref{drmhd}) that the current sheets are not formed in isolation, but are a result of evolution of high-density regions, which are driven together by the evolution of the system.  Therefore, fast reconnection is expected in our simulations. 

Fig.~\ref{mag} in Appendix~\ref{app:plasmoids}  shows selected reconnection layers in the chosen 2D simulations Ideal-MHD and Rel-MHD. When magnetic field lines reconnect, the magnetic tension acts to shorten the field lines and make a magnetic slingshot, which drives the outflow (plasmoids ejection) from both sides of the layer in the parallel direction \citep{10.1007/BFb0102431, linton2001reconnection}. 

For a steady-state reconnection, the outflow (from the reconnection area) should be balanced with the inflow (toward the reconnection layer) which is shown with the white arrows in the figure. The ratio of inflow and outflow velocity ($v_{\rm in}$ and $v_{\rm out}$, respectively), is called the reconnection rate $v_{\rm r}=v_{\rm in}/v_{\rm out}$. 

The outflow propagates along the background magnetic field lines with the Alfv\'{e}n speed $ v_{\rm A} =c\sqrt{\sigma/(\sigma+1)}$, in conventional units. When $\sigma \simeq 10$, $v_{\rm A} \simeq c$, the reconnection rate can be approximated with $v_{\rm r}=v_{\rm in}/c$.
The magnetization values on both sides of the reconnection layer in all simulations are greater than $8$, as demonstrated in Appendix~\ref{app:plasmoids} (Figs.~\ref{3Dyslice}, \ref{3Dzslice} and \ref{mag}).
To compute the reconnection rate we average the inflow velocity of $6$ grid cells located on both sides of the layer. The structure of the layer is found by the Harris equilibrium method \citep{Harris_1962, BR01}. 

According to analytical and numerical studies, the reconnection rate in 3D might be both lower or higher than in 2D. The reconnection rate depends on different parameters such as the initial setup, strength of the magnetic field, and turbulence of the system.
\citet{cemyang14} studied magnetic reconnection in 2D and 3D geometries using resistive MHD simulations and found that the reconnection rate in 3D was approximately twice as fast as in 2D. \citet{Huang_2016} found that in some cases the 3D reconnection rate can be lower than the 2D reconnection rate due to the complex interplay between the plasmoid instability and the turbulent background. 

Our study presents various initial setups in both two and three dimensions (Section \ref{inb}) that affect the magnetization on both sides of the connection region, which in turn influences the reconnection rate. Our Ideal-MHD simulations result in faster reconnection in 3D than in 2D, while the opposite is observed in the Rel-MHD simulations, where the reconnection rate is slower in 3D. In Fig. \ref{vr_R} we show $v_{\rm r}$ as a function of resolution in the simulations with KORAL simulations. We summarize the results of Fig.~\ref{vr_R} as follows.

\begin{figure*} 
\includegraphics[width=1\textwidth]{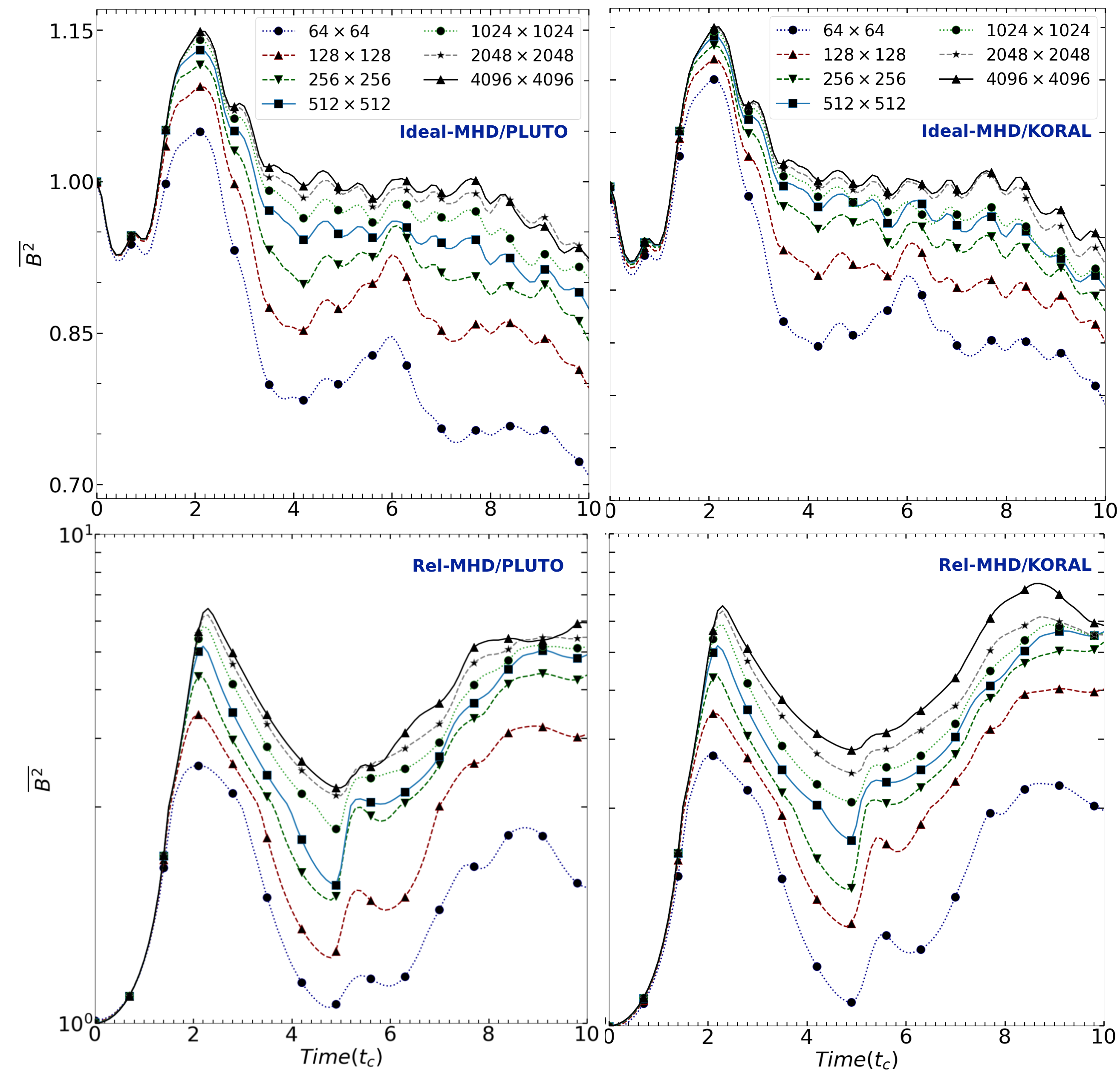}
\caption{Time evolution of $\overline{B^2}$ in simulations with different resolutions using PLUTO ({\it left panels}) and KORAL ({\it right panels}) for the simulations Ideal-MHD ({\it top panels}) and Rel-MHD ({\it bottom panels}). The value of $\overline{B^2}$ is slightly higher in the simulations with KORAL. Note: the y-axis is common between left and right panels, and the legend is the same for all panels.}
\label{pk}
\end{figure*}

In 2D setups:

1) Results of the Ideal-MHD simulations show that the resolution does not affect the reconnection rate in the resolutions $\geq256^2$. We confirm that in the non-relativistic simulations, the current sheet is well resolved in the resolutions $\geq256^2$ (It is also shown in the top panels of Fig. \ref{pk} at $t\simeq 2.5 t_c$ that the curves of $\overline{B^2(t)}$ at higher resolutions are convergent). In the lower resolutions the reconnection rate changes as a function of resolution $v_{\rm r} \approx 0.04 \Tilde{R}^{-0.7}$ ($\Tilde{R} = R/100$). 

2) Results of the Rel-MHD simulations show that the reconnection rate changes as a function of the resolution as $v_{\rm r} \approx 0.25\Tilde{R}^{-0.45}$ in the resolutions $\leq 2048^2$. The current sheets and plasmoids are well resolved in the two highest resolutions. 

In both Ideal-MHD and Rel-MHD simulations in the lowest resolutions ($64^2$ and $128^2$), the numerical resistivity is much higher than $10^{-4}$, and the current layer is not resolved. The reconnection rate converges to a constant value at a lower resolution in the Ideal-MHD than in the Rel-MHD simulations. Therefore, in Rel-MHD, it is necessary to increase the resolution with respect to the non-relativistic case to reach a reconnection rate limit that is resolution independent.

In 3D setups, the current sheets are not resolved in the resolution $128^3$. With the higher resolutions $256^3$ and $512^3$, we do not see a significant effect of the resolution. 
In KORAL the lowest value of reconnection rate in 2D simulations at the highest resolution is about 0.1 in the Ideal-MHD and about 0.16 in the Rel-MHD. In 3D simulations, the value of the reconnection rate in the highest resolutions is around $0.3$ in both Ideal-MHD and Rel-MHD simulations.

Turning to the resisitive simulations, in Fig.~\ref{vr_eta} we plot the reconnection rates of Res-MHD runs with $\eta = 10^{-4}, 10^{-3}$, and $5 \times 10^{-3}$ in the resolution $512^2$ in 2D and $512^3$ in 3D. The reconnection rate changes as a function of resistivity, increasing by a factor of about 60\% in the 3D case and 30\% in the 2D one, as the resistivity changes from $10^{-4}$ to $5\times10^{-3}$. This increase is much smaller than the factor 7.07 expected from the Sweet-Parker law ($v_r\propto \eta^{1/2}$). The dependence seems to be consistent with $1/\log\eta$, instead.
%

Given our fairly low resolution and the small number of points, we cannot make definite claims about the functional form of the reconnection rate. However, the reconnection rate we find is consistent with the dependence on the Lundquist number predicted in Petschek reconnection \citep[$v_r = a/\log S$,][]{Petscheck1964}. The proportionality constant is $a=0.34$ for the 3D simulations,\footnote{For the 3D simulation, $v_r$ is within $\sim10\%$ of $\pi/(8\log_{10} S)$, assuming $v_{\rm A}L=1$. An accurate fit to this formula can be found if we allow values of the characteristic scale $v_{\rm A}L$ to be slightly larger than unity.} and $a=0.10$ for the 2D simulations. Here we assumed $v_{\rm A}L=1$ and we take logarithms to the base 10 ($\log\equiv\log_{10}$). Since our flow is not strongly magnetized nor highly turbulent, the reconnection rate in our resistive simulations is below the rates from \cite{Lazarian99}.

\section{Code comparison} \label{Scodes}
%
The codes we used in our simulations, PLUTO and KORAL, rely on solving the MHD equations (given in Section \S\ref{formalism}) employing the finite volume method. The initial equations are typically formulated in terms of the primitive variables, which include the fluid density, pressure, and velocity, as well as the magnetic field (given in Section \S\ref{inb}).
To solve the equations using the finite volume method, the computational domain is divided into a grid of cells, each of which contains a set of conserved quantities. These conserved quantities are related to the primitive variables through a set of conversion equations, which are typically derived from the conservation laws of mass, momentum, and energy.
Although both PLUTO and KORAL employ the same scheme to calculate conserved quantity fluxes at the boundary of each grid cell, the conversion of primitive to conserved quantities differs between the two codes. PLUTO employs the inversion scheme provided by \citet{m07}, while KORAL uses the $1D_W$ inversion scheme outlined in \citet{Noble_2006}.

We perform simulations of the OT test problem with PLUTO and KORAL codes in the simulations Ideal-MHD and Rel-MHD. The same initial conditions are used in both codes. Here we compare the energy components in the results, the ability of the codes to capture substructures, and the reconnection rates. 
\begin{figure}
\includegraphics[width=1\columnwidth]{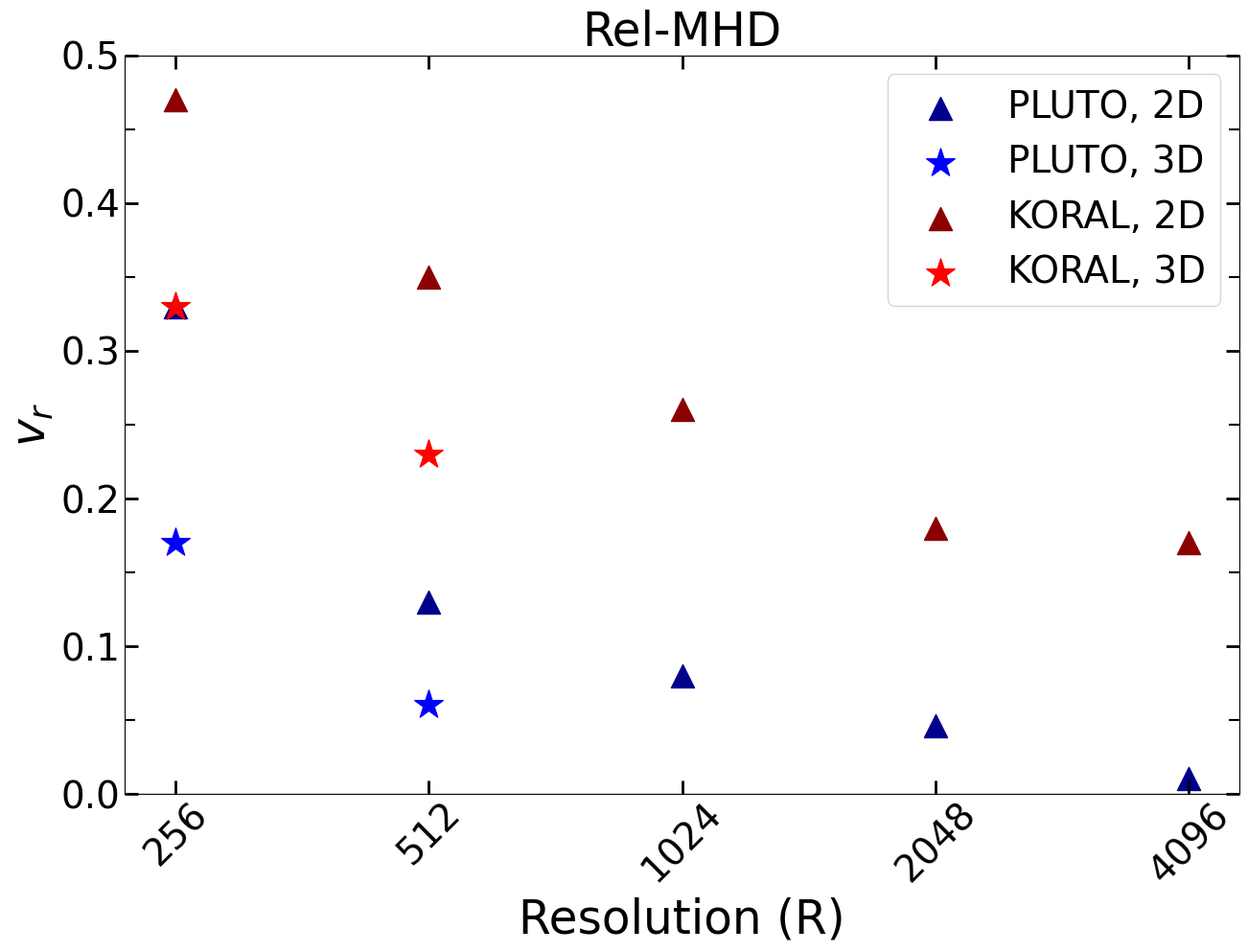}
\caption{The reconnection rate as a function of resolution in simulations Rel-MHD in 2D and 3D. Red symbols indicate simulations with KORAL and blue symbols with PLUTO.}
\label{vr_PK}
\end{figure}
%
In Fig.~\ref{pk} we present the time evolution results for the magnetic energy in the Ideal-MHD and Rel-MHD simulations in PLUTO and KORAL. The value of $\overline{B^2}$ in the simulations Ideal-MHD slightly increases in KORAL with respect to PLUTO. This difference in the value of $\overline{B^2}$ is more obvious in the lower resolutions and in the later time steps. In addition, in Fig.~\ref{totrnr}, we showed that at the identical time steps of Rel-MHD simulation, the residual of the total energy in Rel-MHD in KORAL is typically slightly higher than the one in PLUTO.

To investigate the difference between the codes we plot in  Fig.~\ref{pkresidual} of Appendix~\ref{app:plasmoids} relative differences between KORAL and PLUTO of various quantities. In 
the Ideal-MHD simulations with sufficient resolution for the small numerical resistivity, both PLUTO and KORAL show almost the same numerical dissipation. In the Rel-MHD simulations, the difference between the codes is more pronounced. Also, by comparing the results in Ideal-MHD and Rel-MHD simulations in Fig. \ref{pk}, we find that the numerical resistivity is negligible at the largest resolution $4096^2$ in the Ideal-MHD simulations (the curves of two larger resolutions overlapping) while in the Rel-MHD simulations, one should increase the resolution to obtain a negligible numerical error. 

As mentioned in \S\ref{Sres}, we expect the plasmoid unstable current sheets when there is a hump in $\overline{B^2}$ plot. We show the rest-mass density plot at $t=2.5 t_c$ in the simulation Ideal-MHD and $t=9 t_c$ in the simulation Rel-MHD at the highest resolution $4096^2$ in Appendix~\ref{app:plasmoids}, Figs.~\ref{dmhd} and \ref{drmhd}. These density plots confirm that KORAL is more precise than PLUTO in capturing the substructures.  

We compare the reconnection rate in the simulation Rel-MHD in PLUTO and KORAL in Fig. \ref{vr_PK}. In Fig. \ref{pkresidual}, we show that the residual relative difference between various quantities in the Ideal-MHD simulation is at the level below 1\%, so we only compare $v_r$ in the Rel-MHD simulation.

We observe that in both 2D and 3D setups the reconnection rate in KORAL simulations is higher than in PLUTO simulations. The magnetization on both sides of the reconnection layer directly affects the reconnection rate (which is discussed in \S\ref{vr}), and we showed that in KORAL simulations the magnetic energy (and corresponding magnetization) is higher than in PLUTO simulations. This causes a higher reconnection rate in KORAL simulations compared to PLUTO simulations, as shown in Fig. \ref{vr_PK}.
\section{Summary and Conclusions}\label{concl}
We investigate how the resolution and dimensionality of the simulation setup affect the energy dissipation, substructure formation, and reconnection rate, all of which are critically dependent in astrophysical simulations on the onset of reconnection. We study these effects by performing the Orszag-Tang test problem in the numerical simulation codes PLUTO and KORAL. We perform a quantitative comparison between the results obtained from various setups, including relativistic, non-resistive MHD (Rel-MHD), non-relativistic, non-resistive MHD (Ideal-MHD), non-relativistic, resistive MHD (Res-MHD), in 2D as well as 3D simulations. 

First of all, we estimated the numerical resistivity of the simulations in each resolution to find a sufficient resolution in which we can resolve the substructures and study the energy conversion in our simulations. We used PLUTO code in resistive and non-resistive modes (Res-MHD and Ideal-MHD, respectively) in non-relativistic simulations. We show that the numerical resistivity in the resolution $512^2$ in both 2D and 3D setups is $\eta\approx 10^{-4}$, which is also the limit of the formation of a plasmoid unstable current sheet.

After finding the sufficient resolution for overcoming the effects of numerical resistivity, we study energy conversion in Ideal MHD, Rel-MHD, and Res-MHD simulations. We showed that in Ideal-MHD and Res-MHD simulations magnetic energy converts into internal energy and heats up the plasma.  In Ideal-MHD simulation a part of magnetic energy converts to kinetic energy which accelerates the plasmoids out of the reconnection layer. We also show that in Res-MHD simulations, as expected, the magnetic energy dissipation increases with increasing physical resistivity. In higher resistivity cases, there is a corresponding increase in internal energy.  

In relativistic simulations, Rel-MHD, we find that the relativistic shocks amplify the magnetic field with the magnetic energy $\overline{E_{\mathrm B}}$ increasing by a factor of five at $t=$ 20\% of total simulation time. It is also shown that magnetic energy converts into internal and kinetic energies which amplify the magnetic field for the second time during our simulation. The second increase in magnetic energy at $t=$ 90\% of total simulation time is coincident with the formation of a set of plasmoid unstable current sheets.

We also compare two state-of-the-art codes, PLUTO and KORAL, in both non-relativistic and relativistic simulations. Our findings indicate that in both Ideal-MHD and Rel-MHD simulations, KORAL simulations show higher magnetic energy, $\overline{B^2}$, (implying less magnetic dissipation) compared to PLUTO with the difference more prominent at low resolutions. We show that in resolution $1024^2$, in the Ideal-MHD simulations, the relative difference of relevant quantities in PLUTO and KORAL is less than $10^{-2}$, while in the Rel-MHD simulations, for some quantities the residual reaches $0.1$. In the highest resolution run ($4096^2$), we found that KORAL captures more substructures than PLUTO in both Ideal-MHD and Rel-MHD simulations. 
We show that the reconnection rate in all simulations in KORAL is higher than that in PLUTO---it is caused by higher magnetization in the reconnection layer region in KORAL.  

We study the effect of resolution on the reconnection rate $v_r$ in our simulations. As expected, numerical resistivity influences the reconnection rate. Increasing the resolution leads to a decrease in both numerical dissipation and reconnection rate. In 2D simulations, $v_r$ is initially  a function of scaled resolution ($\Tilde{R}=R/100$) as $v_{\rm r} \approx 0.04 \Tilde{R}^{-0.7}$ (Ideal-MHD) and $v_{\rm r} \approx 0.25\Tilde{R}^{-0.45}$ (Rel-MHD). In each set of simulations, we find a resolution beyond which the reconnection rate is no longer affected by the resolution, and we find the limiting reconnection rate in this limit:
in 2D simulations in KORAL, in the Ideal-MHD runs, $ v_{\rm r} = 0.1$ for resolutions $\geq 512^2$;  in the Rel-MHD, $v_{\rm r} \approx 0.18$ for resolutions $\geq 2048^2$.
In PLUTO simulations, the reconnection rate is lower than that in KORAL simulations. In PLUTO, in Ideal-MHD $v_r \approx 0.03$, in Rel-MHD $v_r \approx 0.05$.

We conclude that the Rel-MHD simulations should be performed at resolutions at least four times larger than in the non-relativistic Ideal-MHD simulations, to reach a negligible effect of the resolution on the reconnection rate.

In 3D simulations in KORAL the Ideal-MHD and Rel-MHD simulations are not directly comparable since we initialized the velocity and magnetic fields differently. Still, in both setups, 
the results are remarkably similar, with the effect of resolution on $v_{\rm r}$  not significant in higher resolutions. In both Ideal-MHD and Rel-MHD simulations with resolution $512^3$ the reconnection rate $v_{\rm r} \simeq 0.3$ (Fig.~\ref{vr_R}). 

When comparing the reconnection rate in 2D and 3D setups, it is crucial to consider several parameters, such as the initial setup, the strength and topology of the magnetic field, and the turbulence of the system. In setups with the equivalent magnetization and turbulence levels, we show that the reconnection rate in 3D ideal MHD simulations is lower than that observed in 2D simulations. This trend is particularly notable in relativistic simulations when comparing the 2D and 3D setups. However, in the resistive runs (Res-MHD) the trend is the opposite, the reconnection rate is about a factor of 3 smaller in 2D simulations than in 3D ones.
We also show that in the resistive simulations, the reconnection rate seems to be well approximated by a $v_r\propto 1/\log\eta$ dependence, reminiscent of Petschek's fast reconnection \citep{Petscheck1964}. 

The results presented here add to the information needed to evaluate the behavior of numerical MHD codes in different setups. The performance of the codes can be evaluated and compared only with a detailed account of the relation between the substructure formation and the amount of energy in each component. By using the standard Orszag-Tang test, we provided detailed quantitative information on energy components, reconnection rates and substructure formation. Our approach can be followed---and the results compared---for other codes.

A caveat in our work here is that, because of the computational expense, we did not follow the convergence of the results in 3D up to the same resolutions as we did in the 2D setups. The new generation of simulations will unavoidably need such an update in benchmarking.  The convergence of vorticity will be addressed in future work.

\section*{Acknowledgements}
 This project was funded by the Polish NCN grant No.2019/33/B/ST9/01564. M\v{C} acknowledges the Czech Science Foundation (GA\v{C}R) grant No.~21-06825X. MW was supported by the European Research Council advanced grant “M2FINDERS - Mapping Magnetic Fields with INterferometry Down to Event hoRizon Scales” (Grant No. 101018682). High-resolution computations in this work were performed on the Prometheus and Ares machines, part of the PLGrid infrastructure. We thank K. Nalewajko and B. Ripperda for inspiring discussions and suggestions. 
\section*{Data availability}
The data underlying this article will be shared on reasonable request
to the corresponding author.
\bibliographystyle{mnras}
\bibliography{otkrefs}

\appendix{
\section{Formation of plasmoids}
\label{app:plasmoids}
In this Appendix, we provide additional figures to support the findings and conclusions presented in the main text. In particular, the plots presented here are intended to help visualizing the plasmoids in 2D and 3D simulations, the reconnection layer and magnetization of the upstream region of the reconnection layer, and the comparison of numerical codes. 
The zoomed panels in Figs. \ref{3Dzslice} and \ref{3Dyslice} correspond to the Ideal-MHD and Rel-MHD simulations, respectively, in 3D, and show a magnetization of $\sigma\approx 10$ on both sides of the reconnection layer.

%
\begin{figure}
\centering
\includegraphics[width=1\columnwidth]{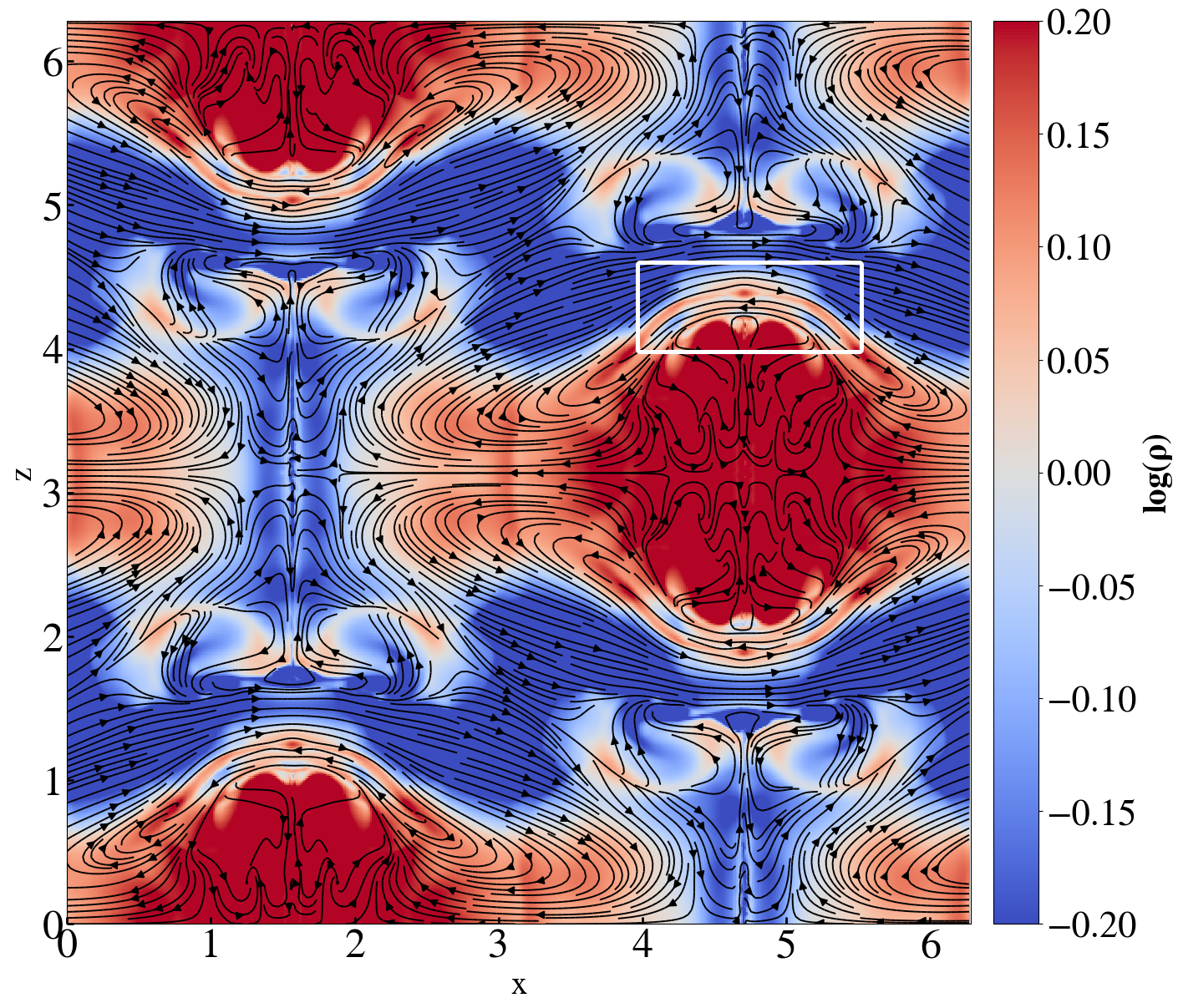}
\includegraphics[width=1\columnwidth]{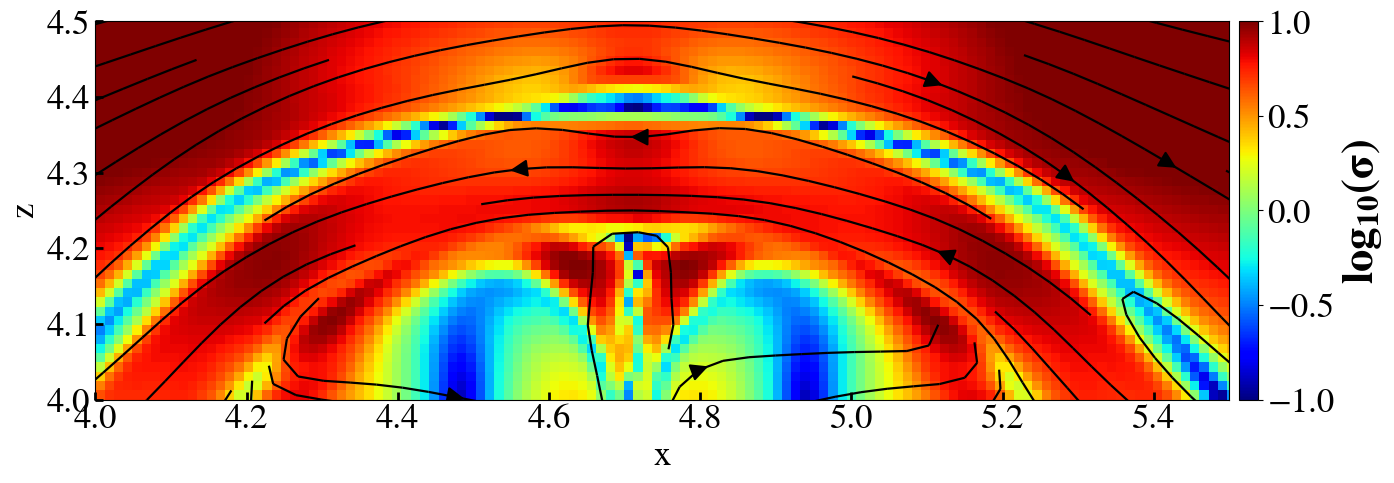}
\caption{In the {\it top panel} is shown a slice in the rest-mass density at $y=3.68$ in the Rel-MHD simulation in 3D with PLUTO, at the resolution $512^3$. The white box shows the reconnection layer contains a plasmoid. The streamlines indicate the magnetic field.  In the {\it bottom panel} we plot the magnetization of the selected region (white box).}
\label{3Dyslice}
\end{figure}

\begin{figure}
\centering
\includegraphics[width=1\columnwidth]{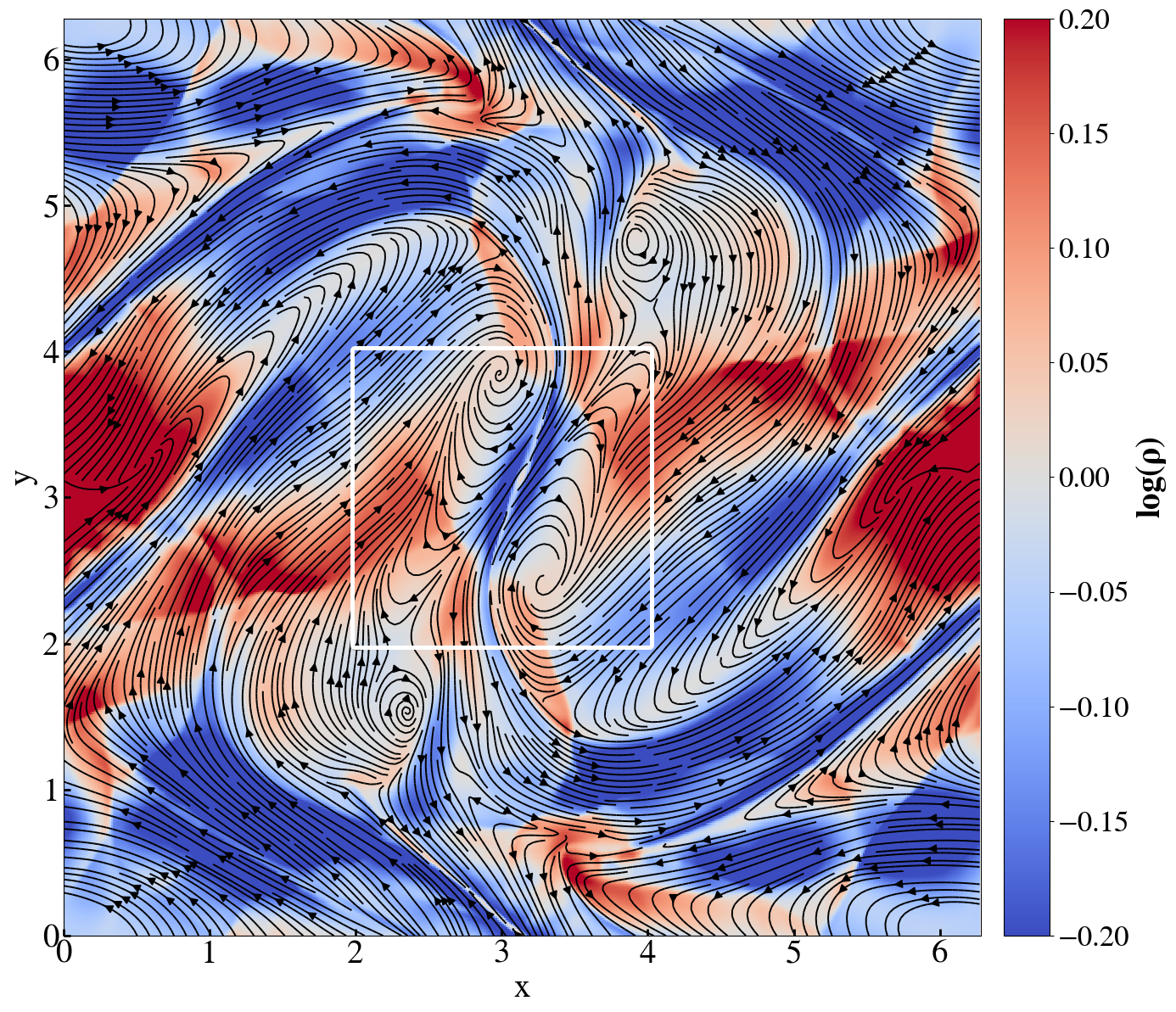}
\includegraphics[width=1\columnwidth]{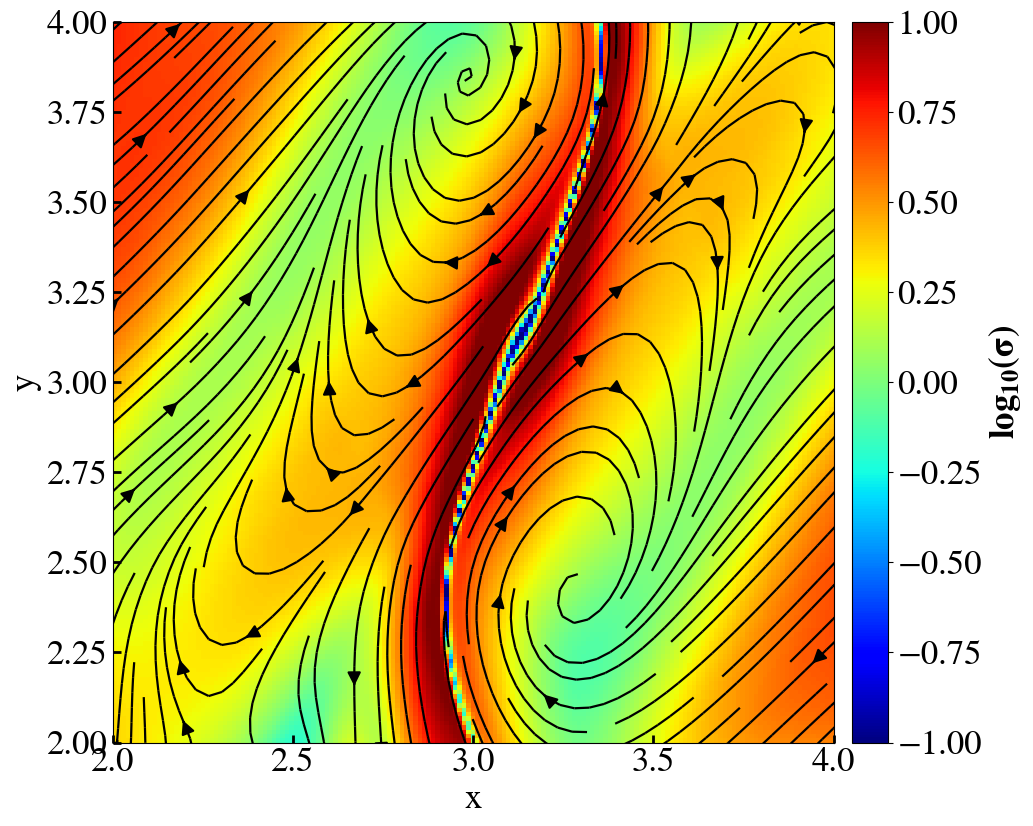}
\caption{In the {\it top panel} is shown a slice in the rest-mass density at $z=\pi$ in the Ideal-MHD simulation in 3D with PLUTO, at the resolution $512^3$. The white box shows the reconnection layer contains a plasmoid. The streamlines indicate the magnetic field.  In the {\it bottom panel} we plot the magnetization of the selected region (white box).}
\label{3Dzslice}
\end{figure}
%
\begin{figure*}
\centering
\includegraphics[width=0.9\columnwidth]{Figs/zMHD_4096_pl_25.png}
\includegraphics[width=0.9\columnwidth]{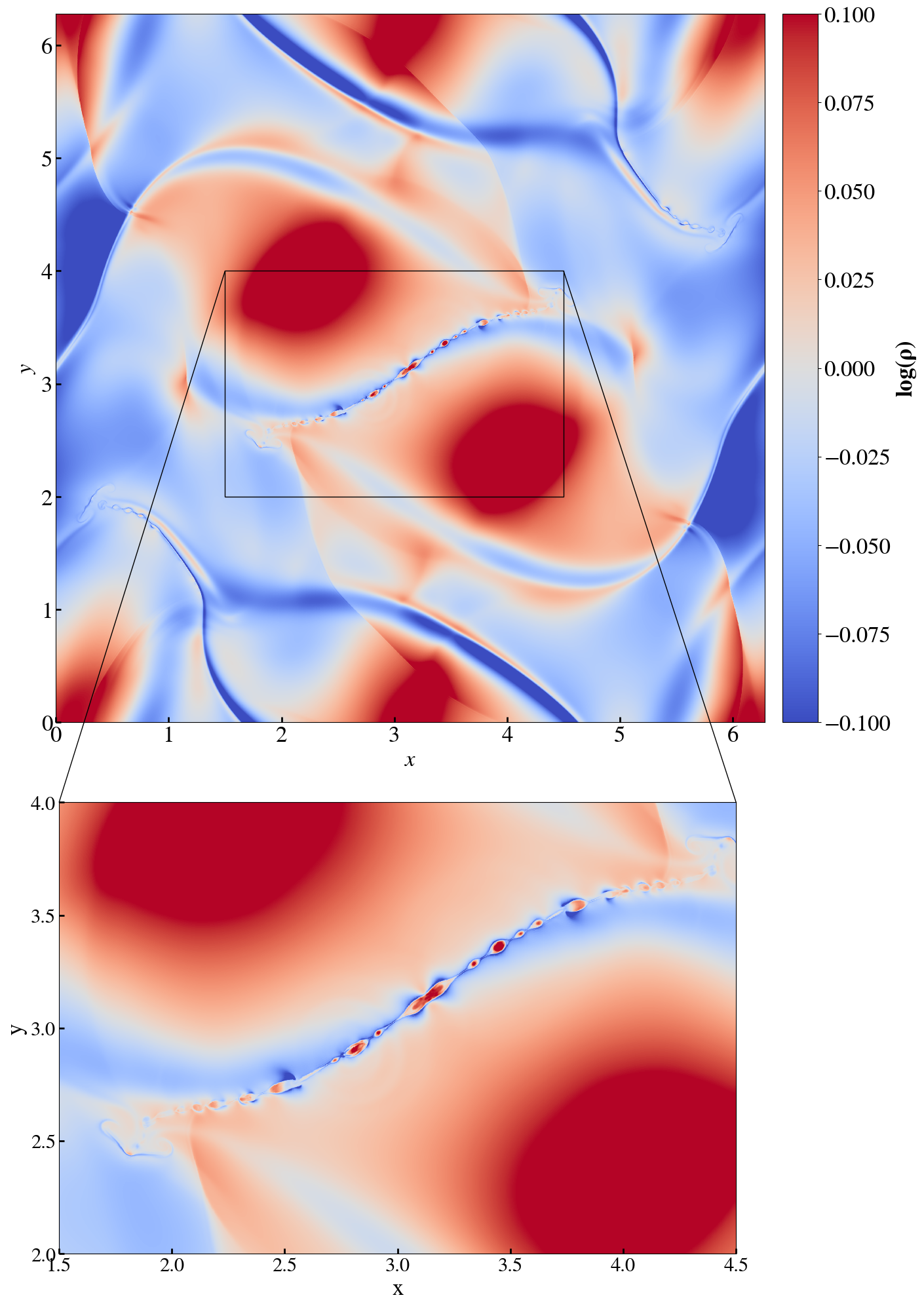}
\caption{The rest-mass density $\rho$ for a vortex at $t = 2.5 t_c$ in the Ideal-MHD simulation at the resolution of $4096^2$ in PLUTO ({\it left panels}) and KORAL ({\it right panels}). Due to lower numerical dissipation, KORAL is more precise in capturing the substructure in the simulations.}
\label{dmhd}
\end{figure*}
%
%
\begin{figure*}
\centering
\includegraphics[width=0.9\columnwidth]{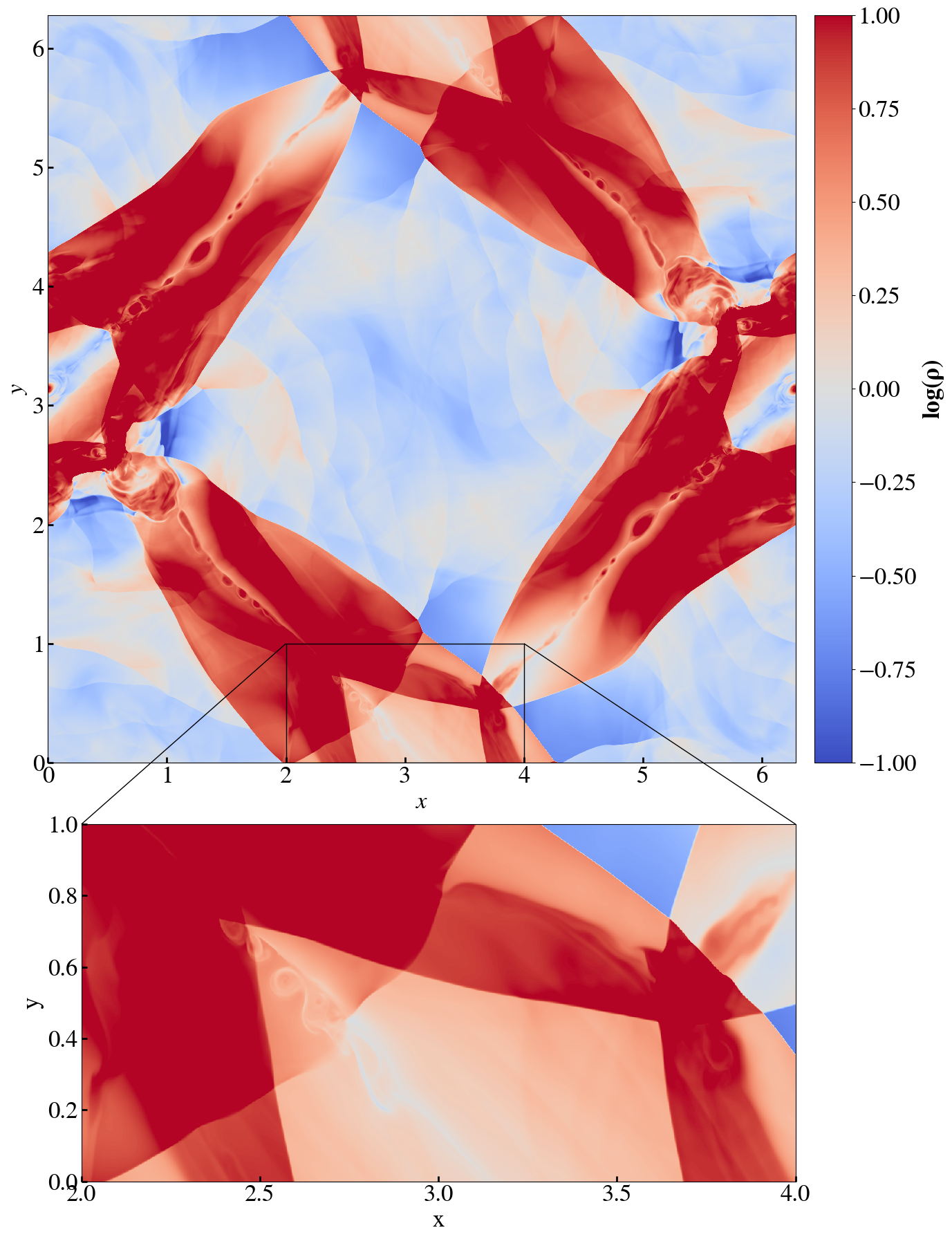}
\includegraphics[width=0.9\columnwidth]{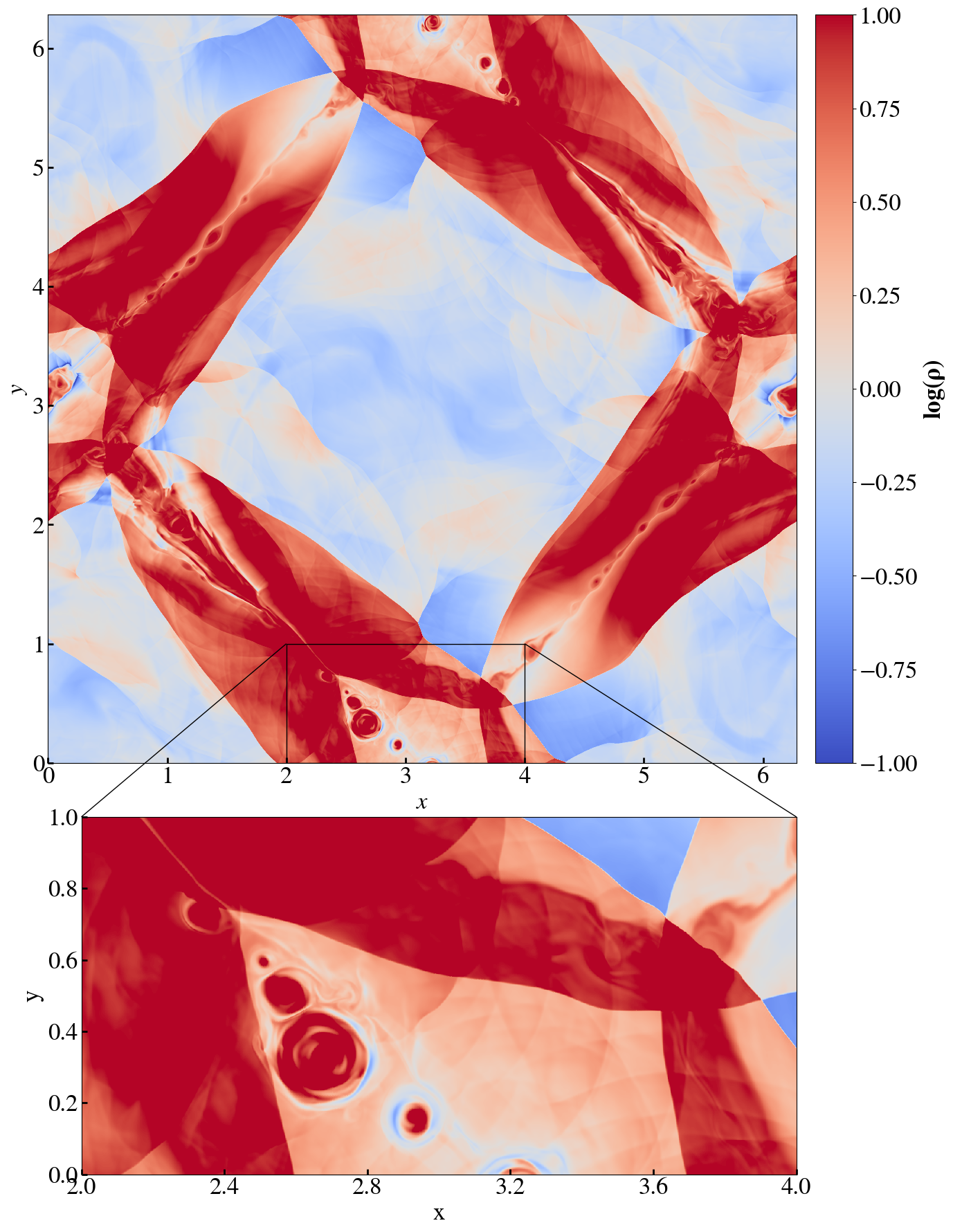}
\caption{Rest-mass density $\rho$ for a vortex at $t = 9 t_c$ of the Rel-MHD simulation with the resolution $4096^2$ in PLUTO ({\it left panels}) and KORAL ({\it right panels}). Due to lower numerical dissipation, KORAL is more precise in capturing the substructure.}
\label{drmhd}
\end{figure*}
%
\begin{figure}
\includegraphics[width=1\columnwidth]{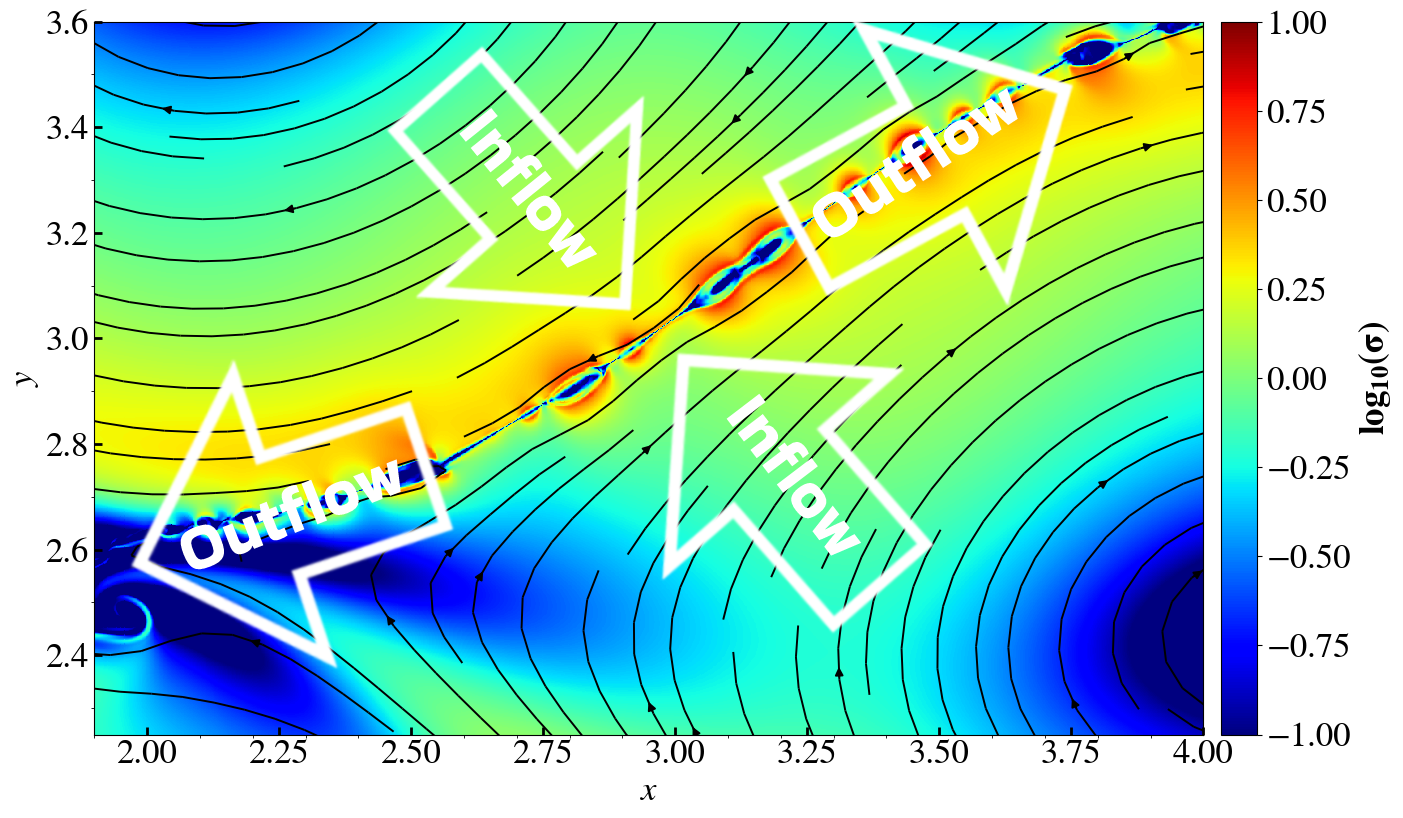}
\includegraphics[width=1\columnwidth]{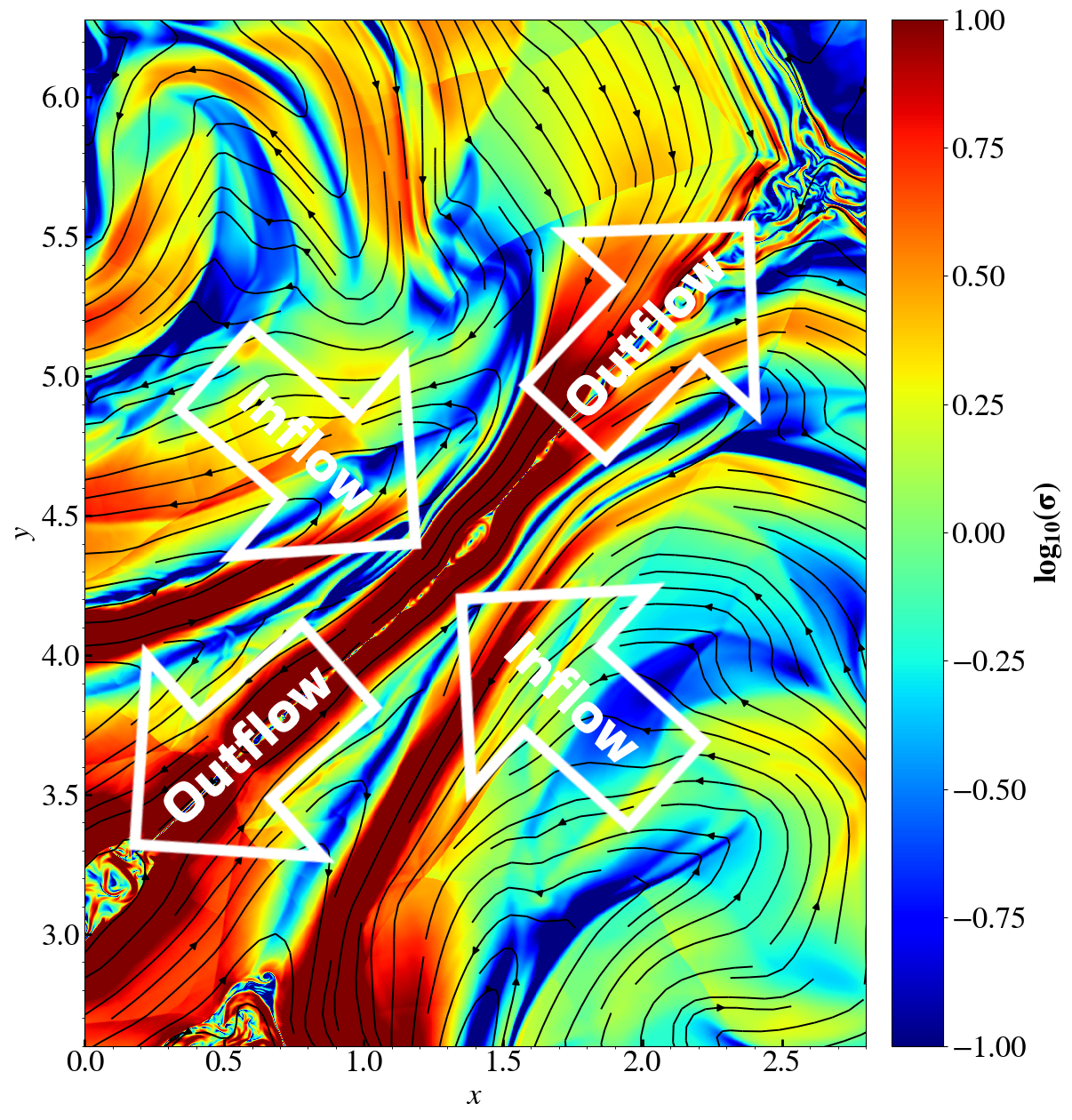}
\caption{The reconnection layer is visible in the magnetization in simulations with KORAL at the resolution $4096^2$ grid cells in the center of the simulation box in the Ideal-MHD simulation ({\it top panel}) and along the ($0,\pi$)-($\pi,2\pi$) line in the simulation box in the Rel-MHD simulation ({\it bottom panel}). The streamlines show the magnetic field.}
\label{mag}
\end{figure}
%
\begin{figure}
\includegraphics[width=1\columnwidth]{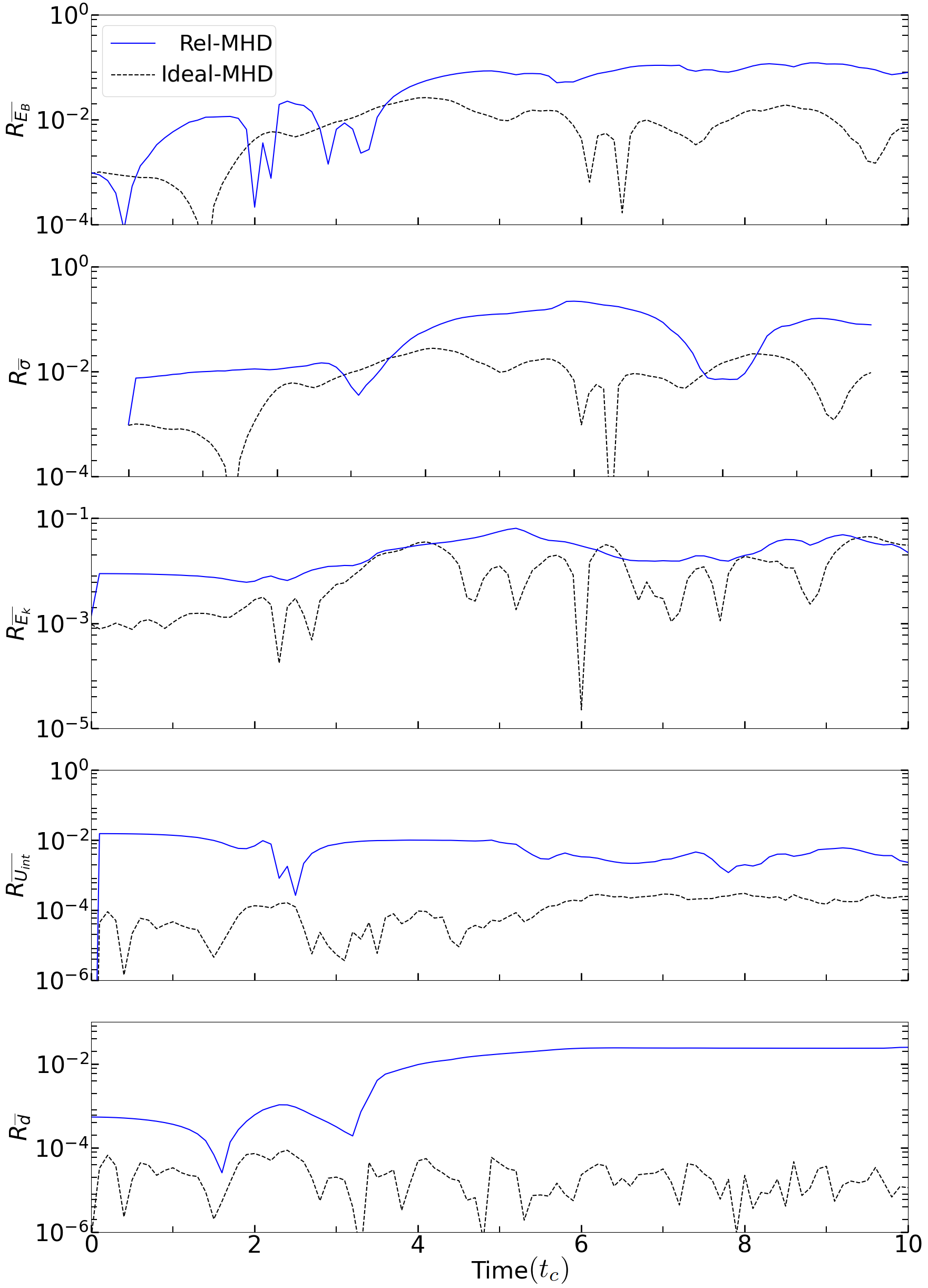}
\caption{The residual of quantities $Q$, $R_Q=|Q_{\rm KORAL}-Q_{\rm PLUTO}|/Q_{\rm KORAL}$, in KORAL and PLUTO with the resolution $1024^2$. We show (from {\sl top} to {\sl bottom}) the residuals for the magnetic energy, $R_{\overline{E_B}}$, magnetization, $R_{\overline{\sigma}}$, kinetic energy, $R_{\overline{E_k}}$, internal energy, $R_{\overline{U_{ \rm int}}}$ and density,  $R_{\overline{d}}$. }
\label{pkresidual}
\end{figure}
%

The rest-mass density in the Ideal-MHD simulations with the resolution $4096^2$ at $t=2.5 t_c$ is shown in Fig. \ref{dmhd}. The zooms in the frames at the bottom panels show the chain of plasmoids in the center of the simulation boxes. In the KORAL simulation, there are more X-points, probably because of the lower numerical resistivity.
The rest-mass density plot of the Rel-MHD simulation with the resolution $4096^2$ is shown in Fig.~\ref{drmhd}. This figure shows the result at a time $t=9t_c$, with the second hump in $\overline{B^2}$ (discussed in \S\ref{Srel}). There are two plasmoid unstable current sheets in the PLUTO simulation box, along the ($0,\pi$)-($\pi,2\pi$) and ($\pi,0$)-($2\pi,\pi$) lines. In the KORAL simulation, two more current sheets are resolved in the bottom and top of the box, along ($\pi,0$)-($0,\pi$) and ($\pi,2\pi$)-($2\pi,\pi$). The zoomed frames at the bottom of this figure show the same regions in PLUTO and KORAL simulation boxes.

Fig. \ref{mag} shows reconnection layers, the so-called magnetic diffusion region, in the Ideal-MHD simulation (top panel) and the Rel-MHD simulation (bottom panel). The color bar indicates the magnetization. The solid lines with arrows represent the streamlines of the magnetic field, pointing in opposite directions around the reconnection layer. The reconnected line (slingshot) can be seen in the plasmoid region at $(0,\pi)$ in the Rel-MHD simulation in the bottom panel. This plot shows that the magnetization $\sigma$ in the upstream region of the current sheet in the Ideal-MHD simulations in 2D is $\approx 8$ and in the Rel-MHD simulations in 2D it is $\approx 10$.

In Fig.~\ref{pkresidual}, we plot the residual quantities
$R_Q=|Q_{\rm KORAL}-Q_{\rm PLUTO}|/Q_{\rm KORAL}$ ($Q$ represents the compared quantity) to clarify the difference between PLUTO and KORAL simulations. The black dashed curves correspond to the Ideal-MHD simulation and the blue solid curves correspond to the Rel-MHD simulation. We compute $R_Q$ in the results with the resolution of $1024^2$, at which the numerical dissipation is small. In the Ideal-MHD simulation, the residuals of magnetic energy ${\overline{E_B}}$ and magnetization ${\overline{\sigma}}$ are less than $10^{-2}$ while in the Rel-MHD simulation, the residuals reach $0.1$. In the  Ideal-MHD simulation, the residual of kinetic energy ${\overline{E_k}}$ is less than $10^{-2}$, while in the Rel-MHD simulation, it is less than $10^{-1}$. The residuals of internal energy ${\overline{U_{ \rm int}}}$ and density ${\overline{d}\equiv\overline{\Gamma\rho}}$ in the Ideal-MHD simulation are of the order of $10^{-4}$ and in the Rel-MHD simulation they are of the order of $10^{-2}$. We conclude 10\% level code consistency for Rel-MHD and 1\% level consistency for non-relativistic Ideal-MHD simulations. 

}
%

\bsp	
\label{lastpage}
\end{document}